\documentclass[fleqn,usenatbib]{mnras}

\usepackage{newtxtext,newtxmath}
\usepackage{booktabs}
\usepackage{threeparttable}

\usepackage[T1]{fontenc}

\DeclareRobustCommand{\VAN}[3]{#2}
\let\VANthebibliography\thebibliography
\def\thebibliography{\DeclareRobustCommand{\VAN}[3]{##3}\VANthebibliography}

\usepackage{graphicx}	
\usepackage{amsmath}	
\usepackage{booktabs}
\usepackage{xcolor}
\usepackage{ulem}
\usepackage{xspace}

\def\sec#1{Section~\ref{sec:#1}}
\def\app#1{Appendix~\ref{sec:#1}}

\def\tab#1{Table~\ref{tab:#1}}

\newcommand{\Chap}[1]{{\protect\hyperref[ch:#1]{Chapter~\ref*{ch:#1}}}}

\newcommand{\Sec}[1]{{\protect\hyperref[sec:#1]{Section~\ref*{sec:#1}}}}

\newcommand{\Fig}[1]{{\protect\hyperref[fig:#1]{Fig.~\ref*{fig:#1}}}}
\newcommand{\Figs}[2]{{\protect\hyperref[fig:#1]{Figs.~\ref*{fig:#1}}~and~\ref{fig:#2}}}
\newcommand{\subFig}[2]{{\protect\hyperref[fig:#1]{Fig.~\ref*{fig:#1}~#2}}}
\newcommand{\Eq}[1]{{\protect\hyperref[eq:#1]{Eq.~\ref*{eq:#1}}}}

\newcommand{\Tab}[1]{{\protect\hyperref[tab:#1]{Table~\ref*{tab:#1}}}}

\newcommand{\App}[1]{{\protect\hyperref[app:#1]{Appendix~\ref*{app:#1}}}}

\newcommand{\ap}{{\scriptsize AUTOPROF }}

\newcommand{\dap}{MaNGA-\textsc{dap}}
\newcommand{\photopaper}{\citetalias{Arora2021}\xspace}

\newcommand{\kms}{\ifmmode\,{\rm km}\,{\rm s}^{-1}\else km$\,$s$^{-1}$\fi}
\newcommand{\Rd}{\ifmmode\,R_{\rm d}\else $R_{\rm d}$\fi}
\newcommand{\be}{\begin{equation}}
\newcommand{\ee}{\end{equation}}
\newcommand\ltsima{$\; \buildrel < \over \sim \;$}
\newcommand\ltsim{\lower.5ex\hbox{\ltsima}}
\newcommand\gtsima{$\; \buildrel > \over \sim \;$}
\newcommand\gtsim{\lower.5ex\hbox{\gtsima}}

\newcommand{\magss}{\ifmmode {{{{\rm mag}~{\rm arcsec}}^{-2}}}
             \else {{{mag}$~${arcsec}$^{-2}$}}
             \fi}
\newcommand{\ha}{H$\alpha$\xspace}

\def \ion#1#2{#1{\footnotesize{#2}}\relax}
\def \hi{\ion{H}{I}\xspace}
\def \littleprime{\ifmmode{\scriptscriptstyle \prime }
     \else{\hbox{$\scriptscriptstyle \prime$ }}\fi}
\def \arcsec{\raise .9ex \hbox{\littleprime\hskip-3pt\littleprime}}

\title[MaNGA galaxy properties -- II]{MaNGA galaxy properties -- II. A detailed comparison of observed and simulated spiral galaxy scaling relations}

\author[Arora et al.]{
Nikhil Arora,$^{1,2,3}$\thanks{E-mail: nikhil.arora@nyu.edu}
Stéphane Courteau,$^{3}$
Connor Stone,$^{3,4,5}$
and Andrea V. Macci\`o$^{1,2,6}$
\\
$^{1}$New York University Abu Dhabi, PO Box 129188, Abu Dhabi, United Arab Emirates\\
$^{2}$Center for Astro, Particle and Planetary Physics (CAP$^3$), New York University Abu Dhabi\\
$^{3}$Department of Physics, Engineering Physics \& Astronomy, Queen's University, Kingston, ON K7L 3N6, Canada\\
$^{4}$Department of Physics, Université de Montréal, Montréal, Québec, Canada\\
$^{5}$Mila -- Québec Artificial Intelligence Institute, Montréal, Québec, Canada\\
$^{6}$Max-Planck-Institut für Astronomie, Königstuhl 17, D-69117 Heidelberg, Germany\\
}

\date{Accepted XXX. Received YYY; in original form ZZZ}

\pubyear{2022}

\begin{document}
\defcitealias{Arora2021}{A21}

\label{firstpage}
\pagerange{\pageref{firstpage}--\pageref{lastpage}}
\maketitle

\begin{abstract}
We present a catalogue of dynamical properties for 2368 late-type galaxies from the MaNGA survey. 
The latter complements the catalogue of photometric properties for the same sample based on deep optical DESI photometry processed with AutoProf.
Rotation curves (RCs), extracted by model fitting H$\alpha$ velocity maps from the MaNGA Data Analysis Pipeline, extend out to 1.4 (1.9)\,R$_{e}$ for the primary (secondary) MaNGA samples. 
The RCs and ancillary MaNGA Pipe3D data products were used to construct various fundamental galaxy scaling relations that are also compared uniformly with similar relations from NIHAO zoom-in simulations.
Simulated NIHAO galaxies were found to broadly reproduce the observed MaNGA galaxy population for $\log (M_*/{\rm M_{\odot}) > 8.5}$.
Some discrepancies remain, such as those pertaining to central stellar densities and the diversity of RCs due to strong feedback schemes.
Also presented are spatially-resolved scatters for the velocity-size-stellar mass (VRM$_*$) structural relations using MaNGA and NIHAO samples.
The scatter for these relations in the galaxian interiors is a consequence of the diversity of inner RC shapes, while scatter in the outskirts is dictated by the large range of stellar surface densities which itself is driven by sporadic star formation. 
The detailed spatially-resolved scatter analysis highlights the complex interplay between local and global astrophysical processes and provides a strong constraint to numerical simulations.
\end{abstract}



\section{Introduction} \label{sec:intro}

Given their importance for constraining galaxy formation and evolution models, studies of galaxy scaling relations have enjoyed a rich history \citep{Faber1976,Tully1977, Bender1992, Mo1998, Steinmetz1999, Courteau2007, GFE2010, Kormendy2013, Lelli2017, Stone2020, Donofrio2021}.
The slope, zero-point, and scatter of scaling relations \citep{Courteau2007, Kormendy2013, Lange2015} encode critical information about the structure of galaxies and provide stringent constraints to galaxy formation models \citep{Dutton2007,Brook2012,Koch2017,Dutton2017, Sande2019, Starkenburg2019}. 
Such data-model comparisons require large unbiased multi-band observed galaxy data \citep{Jarrett2000, York2000, gama} and modern state-of-the-art galaxy formation simulations \citep[see][for a review]{Vogelsberger2020}.

The advent of galaxy surveys with large scale integral field spectroscopy such as MaNGA \citep{Bundy2015, Wake2017}, SAMI \citep{sami}, ATLAS$^{\rm 3D}$ \citep{atlas3d} and CALIFA \citep{Walcher2014} has opened up new areas of investigations with high quality spatially-resolved dynamical and chemical properties of galaxies. 
For instance, the coupling of IFU data with deep multi-band photometric observations \citep[][hereafter \photopaper]{Courteau1996, Hall2012, Gilhuly2018, Ouellette2017, Arora2021} allows detailed studies of different astrophysical processes such as star formation \citep{Pandya2017}, baryonic feedback, dynamics, interactions between baryon and dark matter \citep{Dutton2011b}, impact of environment \citep{Peng2012}, chemical evolution \citep{Gallazzi2005, Menguiano2020}, angular momentum distributions \citep{Romanowsky2012,Obreschkow2014}, and more. 
On the theoretical front, modern state-of-the-art cosmological simulations in a cosmological cold dark matter paradigm allow tracking of the co-evolution of baryons and dark matter \citep{nihao_main, Crain2015, Remus2017, Pillepich2018, Hopkins2018, Habouzit2019}.
While the distribution (but not the nature) of dark matter in the Universe is fairly well understood, especially on large-scales, modelling stars, their formation, and gas distributions within galaxies still challenges most galaxy formation and evolution theories \citep{Avila-Reese2011, Sawala2011, Weinmann2012, Agertz2015}.
Modern large-scale surveys such as the ones mentioned above provide rich and versatile data that can be used to constrain the distribution of various baryonic properties in cosmological simulations. 

Providing the most recent, up-to date and uniform comparison between simulations and observations of galaxies is the motivation for this study.
Such comparisons highlight not only benchmarks for our current understanding of galaxy formation and evolution, they also identify specific areas of improvement for galaxy formation models. 
Different approaches can be taken to enable uniform data-model comparisons of galaxies.
Cosmological simulations can be post-processed to generate mock images and spectra of galaxies that include observational characteristics such as atmosphere blur, background sky noise, photometric bandpasses, etc \citep{Torrey2015, Bottrell2017, Elagali2018, Deeley2021, Bottrell2022, Camps2022}. 
These can then be compared directly with observations. 
Conversely, all known biases and sources of error from observations (such as extragalactic and Galactic dust extinction, photometric band effects, inclination effects, sky noise, sample completion, etc.) can be modeled and removed from the inferred data for direct comparison with intrinsic properties of simulated galaxies \citep{arora2019, Stone2021, Frosst2022}.
The consistency of these different data-model comparisons can provide an internal validation of our methods and results \citep{Stoppa2023}.

Combining the MaNGA photometric catalogue (\photopaper) and auxiliary data from Pipe3D \citep{Sanchez2018} with robust rotation curves yields an extensive catalogue of inferred galaxy structural data which is ideal for comparison with galaxy formation models.
For this paper, the latter is based on the Numerical Investigation of a Hundred Astrophysical Object (NIHAO) project which provides $\sim$60 zoom-in late-type galaxy simulations. 
Data-model comparisons of galaxies often focus on various informative properties of galaxies such as star formation rates \citep{Starkenburg2019}, galaxy sizes \citep{Graaff2022}, shapes of rotation curves \citep{Santos2018}, black hole scaling laws \citep{Fire_BH2022}, etc.
With the substantial MaNGA data presented in \photopaper and here, we can create 12 galaxy scaling relations out of 7 galaxy structural parameters.
The multiple scaling relations \citep{Dutton2011,Trujillo-Gomez2011,Brook2012} allow for a multi-dimensional data-model comparison that identifies specific parameters which simulations struggle to reproduce.

Another important aspect of observation-simulation comparisons is the quantification of fit parameters of scaling relations, such as their slope and scatter.
While slope comparisons between simulations and observations are relatively trivial, scatter comparisons require the removal of observed biases (distance uncertainty, disk thickness, inclination, mass-to-light conversions, etc.)
A common approach to handling such biases consists of removing observational errors in quadrature to retrieve the intrinsic scatter of galaxy scaling relations \citep{Strauss1995, Saintonge2011, Hall2012, Lelli2017}.
However, such a method ignores the correlations between the different biases leading to an underestimation of intrinsic scatter. 
A Bayesian formalism to calculate intrinsic scatter \citep{Stone2020} takes correlated observed errors into account and returns robust estimates of intrinsic scatter.
The Bayesian intrinsic scatters for scaling relations determined here can serve as an accurate robust test for galaxy formation simulations.

Modern deep multi-band imaging and large IFU surveys also allow the study of spatially-resolved scaling relations (e.g., Radial Acceleration Relation \citep{Lelli2017, Stone2019} and the Star Formation Main Sequence \citep[hereafter ``SFMS'';][]{Cano2016, Hall2018}).
Recent studies of spatially-resolved scaling relations have already revealed new aspects of galaxy formation and evolution.
For example, the presence of a tight spatially-resolved SFMS is comparable to the global SFMS \citep{Wutys2013, Delgado2016, Wang2017, Hall2018, Ellison2018}.
The existence of a spatially-resolved star formation law requires a strongly dependent on small-scale local gas and stellar surface density.

Spatially-resolved scaling relations based on a wide array of structural parameters are still largely lacking.
The variations of slope and scatter as a function of galactocentric radius for scaling relations (such as size-mass, size-velocity, and Tully-Fisher relations) can be a powerful tool in identifying the drivers of diversity in galaxy structure.
Spatially-resolved scatter diagnostics can simultaneously constrain the overall distribution of galaxy properties 
and inform us about more advanced tests for cosmological simulations (especially pertaining to astrophysical processes on sub-galactic scales).
With our deep photometric and dynamical data, for the first time, we study the variations and drivers of scatter as a function of location in observed and simulated galaxies. 

This paper is organized as follows: \sec{data} describes the photometry and rotation curves (RCs) from the MaNGA galaxy survey and the process of extracting and correcting various galaxy properties. 
We also compare our RCs against other observed rotation curves and scaling relations. 
\sec{nihao} gives a brief description of the NIHAO galaxy formation simulations that are compared against observations.
We begin our data-model comparison in \sec{one_d} to show one dimensional comparisons of various galaxy properties.
A simultaneous comparison of twelve major galaxy scaling relations between MaNGA and NIHAO, including their observed Bayesian intrinsic scatter, is then presented in \sec{sr}.
In \sec{char_radius}, we examine the radial variations of the spatially-resolved slopes and scatters for structural galaxy scaling relations for MaNGA and NIHAO.
Concluding remarks are offered in \sec{conclusions}, with an eye towards improving numerical simulations of galaxy formation and evolution.
All of our processed data, including the observed light profiles (in \photopaper) and RCs (\app{cat}) are publicly available.

\section{Observational Data} \label{sec:data}

In this section, we briefly describe our extraction of photometric properties for the MaNGA galaxies \citep{Bundy2015, Law2016, Wake2017} using deep imaging from the DESI survey. 
A detailed description of the extensive imaging catalogue is found in \photopaper.
RCs rotation curves and dynamical properties are also derived below, and compared 
against other published rotation curves and scaling relations. 

\subsection{Photometry}
\begin{figure*}
    \centering
    \includegraphics[width=\linewidth]{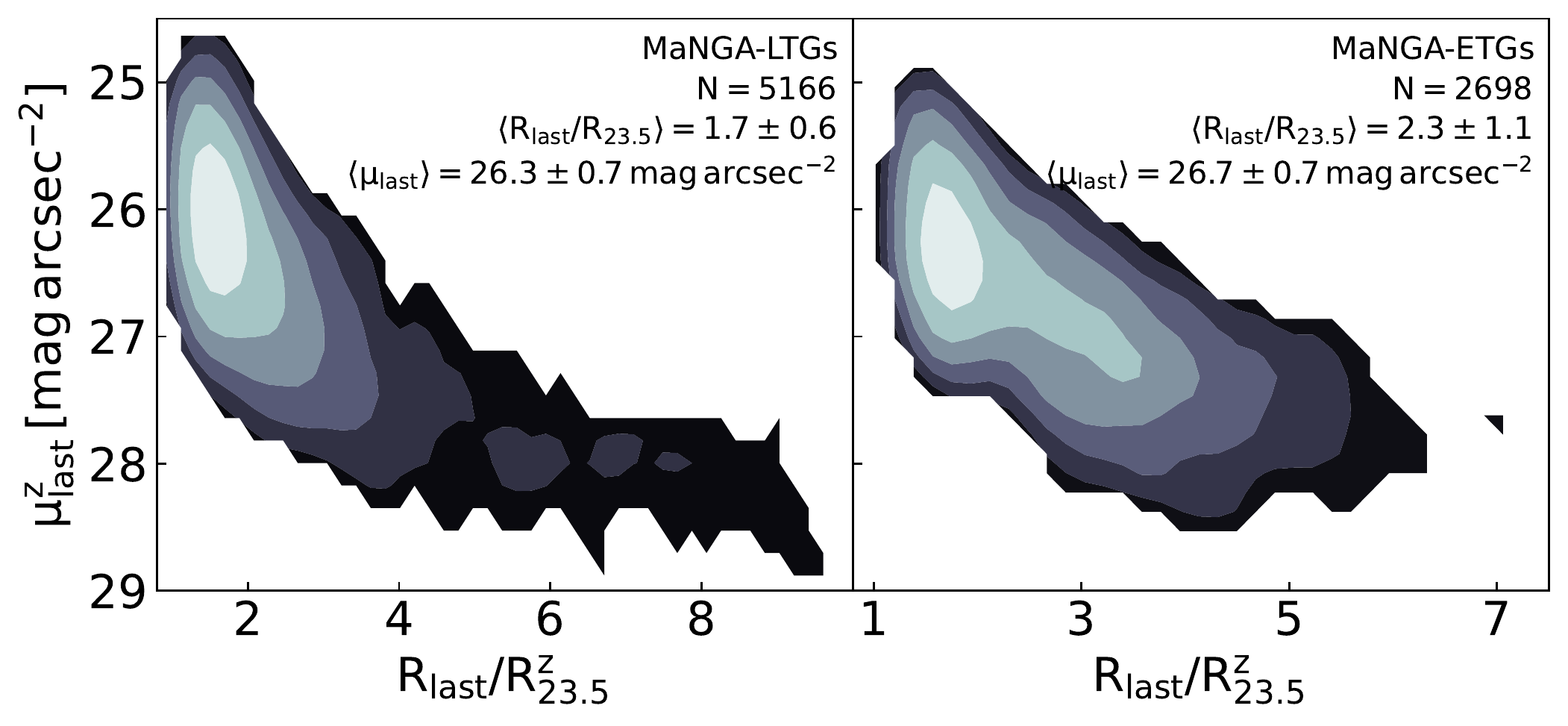}
    \caption{
     Surface brightness depth versus maximal spatial extent (in units of R$_{\rm 23.5}$) for MaNGA galaxies. 
     The surface brightness profiles for LTG (left) and ETGs (right) were extracted through \textsc{autoprof} in the $z$-band.
     The text insets give the number of galaxies, mean maximal spatial extent, and mean surface brightness depth in the $z$-band.}
    \label{fig:sbphoto_depth}
\end{figure*}

In order to lay the foundations for our study of the photometric and dynamical properties of MaNGA galaxies, we first take advantage of the extensive catalogue of non-parametric multi-band photometric and environmental properties for $\sim$4500 MaNGA galaxies, as presented in \photopaper.
This photometric catalogue relied on the deep optical \textit{grz}-band imaging from the Dark Energy Sky Instrument Legacy Imaging Survey \footnote{\url{https://www.legacysurvey.org}} \citep[hereafter DESI]{desi, Dey2019} and the WISE Large Galaxy Atlas and the Extended Source catalogue (WXSC; \citealt{Jarrett2019}).
Non-parametric azimuthally-averaged surface brightness (SB) profiles were extracted using the automated astronomical image analysis tool, \textsc{autoprof} \citep{Stone2021}.
\textsc{autoprof} performs a robust background subtraction and finds an accurate galaxian center for each galaxy image before fitting elliptical isophotes.  

The center, position angle (PA), and ellipticities of each isophote is fit to the DESI \textit{r}-band images for MaNGA galaxies.
The \textit{r}-band imaging is chosen for the extraction of isophotal profiles due to its high signal-to-noise ratio and low dust extinction relative to the other bands (\photopaper).
To obtain multi-band photometry, the \textit{r}-band isophotes are applied to the other \textit{gz}-band images via forced photometry, resulting in complimentary \textit{grz} surface brightness profiles.
The ``forced photometry'' component of \textsc{autoprof} ensures uniformity in measurement of surface brightness, ellipticities, fluxes, and color gradients.
For a more comprehensive description of the \textsc{autoprof} software package and its capabilities, the reader is referred to \cite{Stone2021}.

Because of their lowest sensitivity to dust extinction, all photometric properties were calculated using the DESI \textit{z}-band surface brightness profiles.
This also ensures uniformity with other studies \citep{Stone2020, Arora2021, Frosst2022, Stone2022}.

While the photometric data presented in \photopaper were based on the galaxies selected using the public SDSS-DR16 catalogue \citep{sdssdr16}, the MaNGA photometric data used in this study are based on the SDSS-DR17 \citep{sdssdr17}.
The cross-correlation between MaNGA data release from SDSS-DR17 and the DESI sky survey yielded 7864 galaxies; 5166 of which were classified as LTGs \citep{mldl}, and used in our comparison between simulations and observations.
\Fig{sbphoto_depth} presents the maximal SB depth in the \textit{z}-band of our photometric LTG (left) and ETG (right) samples. 
Thanks to the deep DESI imaging and the versatility of \textsc{autoprof}, we can probe SBs down to (on average) $\sim$26.3\,mag\,arcsec$^{-2}$ and $\sim$2\,R/R$_{23.5}^{z}$ for MaNGA LTGs.
Our robust non-parametric photometry through DESI imaging and \textsc{autoprof} results in gathering 0.1 dex more light and 0.3 dex more stellar mass for the same galaxies than presented in the NSA catalogue as shown in \photopaper. 

Any differences between the photometric data presented here and in \photopaper is due to sample size (using the updated SDSS-DR17). 
The method of extracting and analysing surface brightness profiles is the same in both studies.

\subsection{Rotation Curves}

\begin{figure*}
    \centering
    \includegraphics[width=\linewidth]{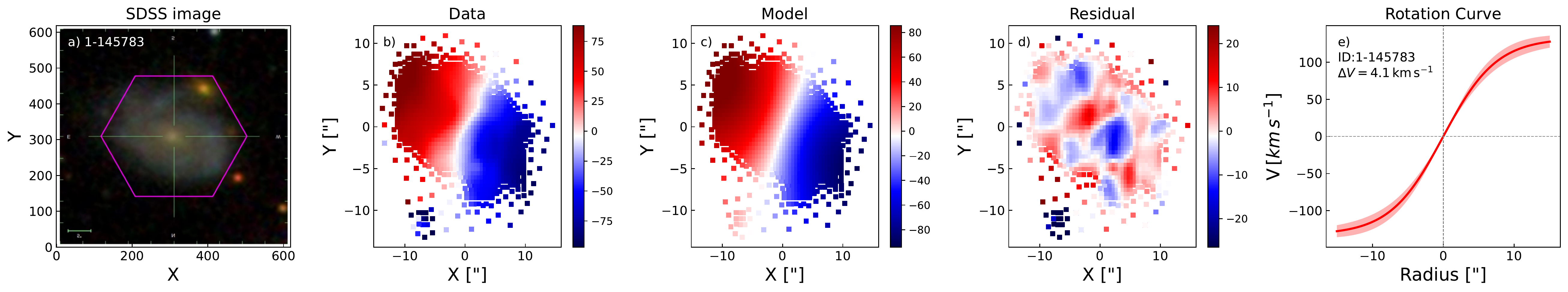}
    \includegraphics[width=\linewidth]{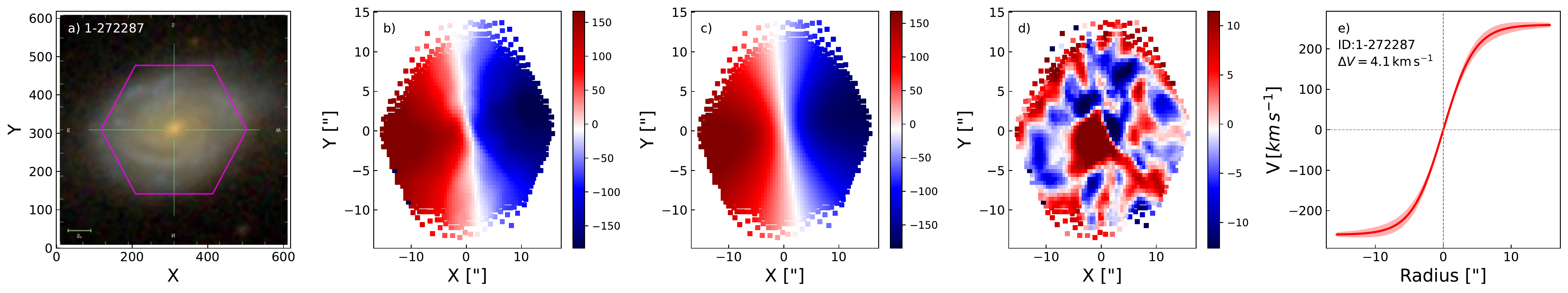}
    \includegraphics[width=\linewidth]{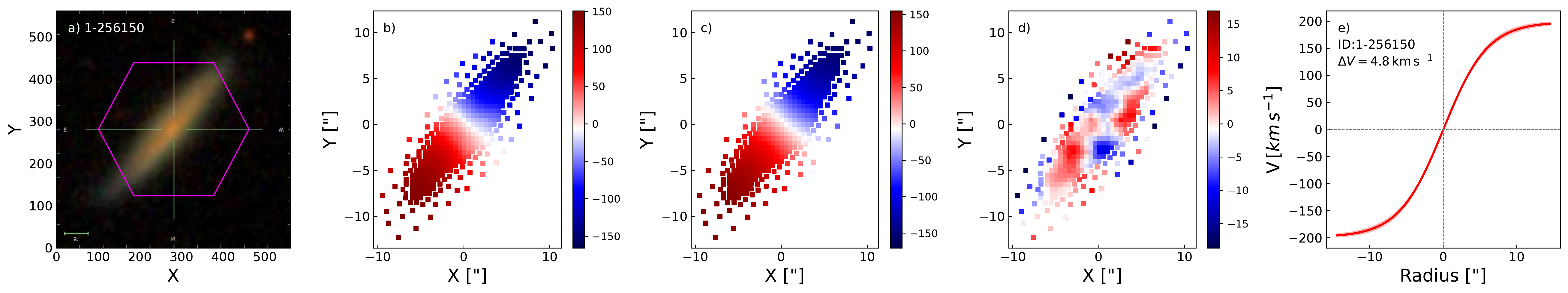}
    \includegraphics[width=\linewidth]{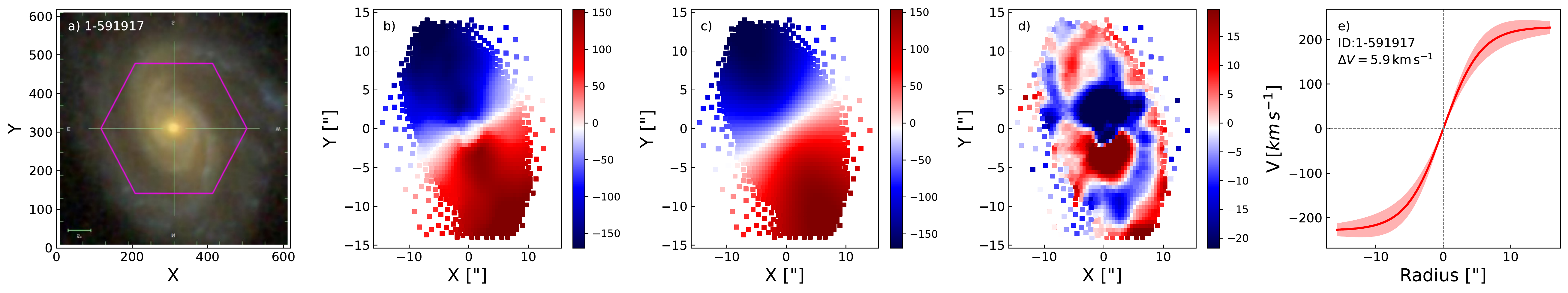}
    \includegraphics[width=\linewidth]{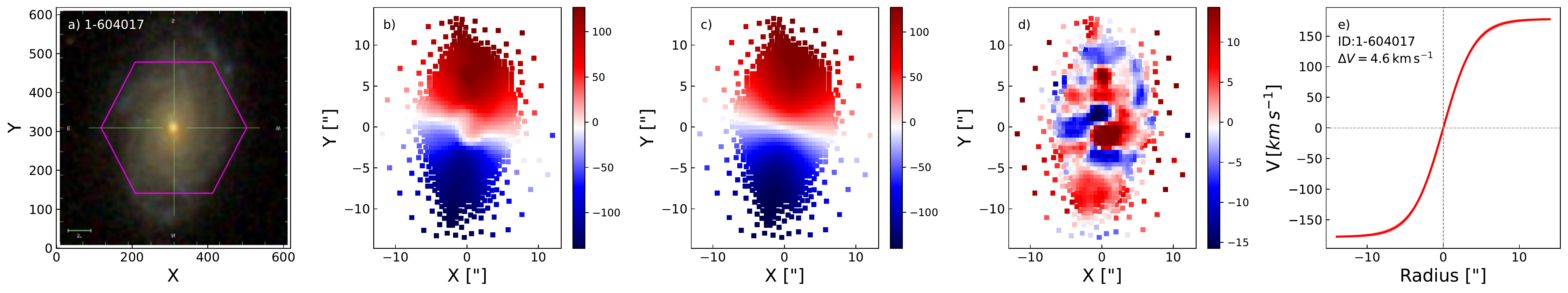}
    \includegraphics[width=\linewidth]{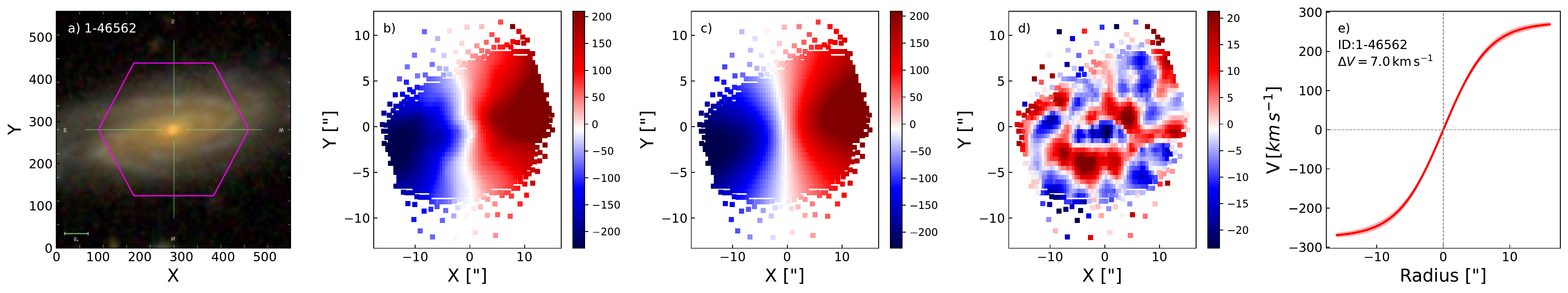}
    \caption{MaNGA \ha velocity maps and extracted RCs. 
    For each row, corresponding to a different MaNGA target, 
    panel a) shows the composite SDSS-\textit{gri} image with the MaNGA IFU footprint (in magenta); panel b) shows the matching \ha velocity map from the MaNGA IFU; and panels c), d) and e) present the tanh model velocity maps, residual maps ($V_{\rm obs}-V_{\rm model}$), and the corresponding rotation curve. 
    The inset in panel e) also gives the MaNGA-ID and average data-model residual $\Delta V\equiv\sqrt{\langle \delta V^2\rangle}$ (see \Eq{verr}).}
    \label{fig:manga_sample}
\end{figure*}

A key aspect of this study is the coupling of kinematic data, presented here, with the photometric information from \photopaper for the same galaxies.
Our spectroscopic analysis takes advantage of the \ha emission line-of-sight (LOS) velocity maps (hereafter \ha velocity maps) provided through the MaNGA Data Analysis Pipeline \citep[\dap;][]{Westfall2019, Belfiore2018} and accessed via the ``Marvin'' toolkit \citep{marvin} for the MaNGA-DR17.
Voronoi binned data are used for the \ha emission LOS velocity maps.

\subsubsection{Velocity model}

RCs were obtained by fitting the following model to the \ha velocity map
\begin{equation}
    V_{\rm LOS}(R, \theta) = V_{\rm sys} + V(R)\cos(\theta)\sin(i),
    \label{eq:vel_model}
\end{equation}
\noindent where $V_{\rm LOS}(R)$ is the observed LOS velocity at radius R, $V_{\rm sys}$ is the systematic (heliocentric) velocity, $\theta$ is the azimuthal angle relative to the positions angle, $i$ is the inclination of the galaxy on the sky, and $V(R)$ is the adopted rotational velocity model.
In \Eq{vel_model}, $R=\sqrt{x_{\rm g}^2 + y_{\rm g}^2}$ and  $\theta=\tan^{-1}(y_{\rm g}/x_{\rm g})$ are calculated using Cartesian coordinates within the disk of the galaxy. 
These are defined as:
\begin{equation}
    x_{\rm g}=x'_{\rm s}\cos(-{\rm PA})+y'_{\rm s}\sin(-{\rm PA}),
    \label{eq:gal_x}
\end{equation}
\noindent and 
\begin{equation}
    y_{\rm g}=\frac{y'_{\rm s}\cos(-{\rm PA}) - x'_{\rm s}\sin(-{\rm PA})}{\cos(i)},
    \label{eq:gal_y}
\end{equation}

where $x'_{\rm s} = x_{\rm s} - x_{\rm c}$ and $y'_{\rm s}= y_{\rm s} - y_{\rm c}$ are centred projected coordinates on the sky with the observed galaxy center given by $(x_{\rm c}, y_{\rm c})$, and PA is the observed position angle.
For this study, and for uniformity with other studies \citep{Courteau1997, Ortiz2020, Brownson2022},
we write:
\begin{equation}
    V(R) = V_{\rm max}\times\tanh(R/R_{\rm t}),
    \label{eq:V_tanh}
\end{equation}
\noindent for the rotational velocity model in \Eq{vel_model}, 
where $V_{\rm max}$ is the maximum velocity, and $R_{\rm t}$ is a turnover radius between the rising and flat portion of the RC. 
Therefore, the kinematic modeling of MaNGA galaxies requires fitting a dynamical model with seven parameters; $(x_{\rm c}, y_{\rm c})$, $i$, PA, $V_{\rm sys}$, $V_{\rm max}$, and $R_{\rm t}$.
The fit parameters are presented in \app{cat}. 

For the fit initialization, $(x_{\rm c}, y_{\rm c})$, and $V_{\rm sys}$ are first set to zero, while the inclination and PA use values calculated through the DESI photometry (\photopaper).
$V_{\rm max}$, and $R_{\rm t}$ are initialized by fitting \Eq{V_tanh} to a ``mock rotation curve'',  calculated by interpolating observed \ha velocity map along to the photometric major axis.
These initialized values are then run through an optimization routine to find the best fit values for the rotation curves.
The quality of the dynamical model is evaluated as the ``mean square error'', 
\begin{equation}
    \langle\delta V^2\rangle = \langle (V_{\rm obs}-V_{\rm model})^2\rangle.
    \label{eq:verr}
\end{equation}.
We calculate robust fit parameters of our dynamical model by running our fitting algorithm 20 times with initialization values randomly chosen within $\pm50$\,per\,cent of the first fit evaluated with the initial values mentioned above. 
The final fit is calculate with the set of fit parameters which return the lowest $\langle\delta V^2\rangle$ out of the random 20 runs.
\Fig{manga_sample} shows the composite SDSS \textit{gri} image, the corresponding \ha velocity map, the model velocity map, the residual and the rotation curves for six late-type MaNGA galaxies.
The top row of \Fig{manga_sample} shows a galaxy with a foreground object; 
in general, we find that our fitting routine is robust against background/foreground objects within the MaNGA IFU.

\begin{figure}
    \centering
    \includegraphics[width=\linewidth]{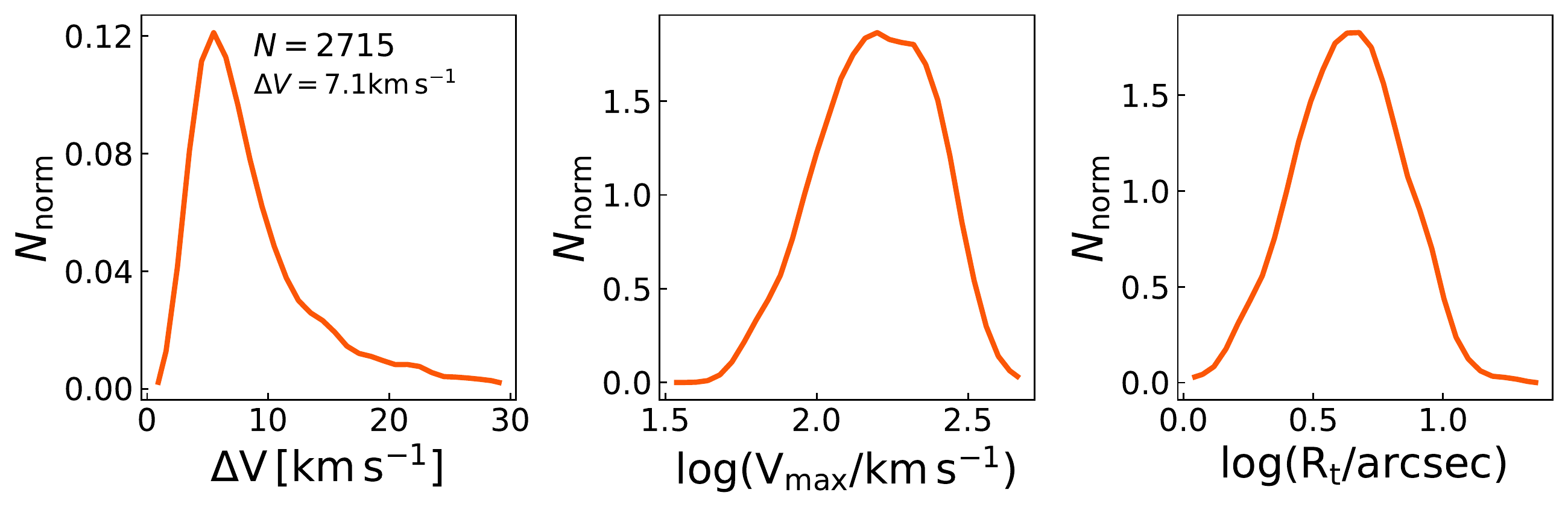}
    \caption{Probability Density Functions normalized using area for the fit quality, $\Delta V$ (see \Eq{verr}; {\it left} panel), as well as the model maximum (asymptotic) velocity, $V_{\rm max}$ ({\it middle}), and turnover radius, $R_{\rm t}$ ({\it right}).}
    \label{fig:mse}
\end{figure}

Our dynamical models fit the observed MaNGA \ha velocity maps very well.
The average $\langle\delta V \rangle$ for all the MaNGA galaxies is 7\,km\,s$^{-1}$ (see \Fig{mse}).
However, the distribution of $\langle\delta V \rangle$ has a non-negligible tail at high $\langle\delta V \rangle$.
Indeed, objects with $\langle\delta V \rangle\geq 10\,{\rm km\,s^{-1}}$ present signatures of non-circular motions (caused by bars and bulges)
that are unaccounted for within our dynamical model.
These non-circular motions (caused by e.g., bars and bulges) are mainly dominant in the central parts of galaxies and have an amplitude of $\sim$15\,km\,$^{-1}$.
Given the stellar mass distribution of MaNGA galaxies ($\log (M_*/\,{\rm M_{\odot}) > 8.5}$); such non-circular motions would not make a large impact on the results presented here.
Given that the galaxy scaling relations that we primarily study trace the outer galaxy regions, the central non-circular motions should be insignificant in our analysis~\citep{Sellwood2021}.

Furthermore, a slight trend is observed between photometric and kinematic inclinations as a function of inclination. 
For more edge-on systems, this offset is $\sim 5-7^{\circ}$. 
This is a result of the strong degeneracy between the V$_{max}$ and the inclination in the \textit{tanh} model.
However, the inclination offset is negligible for tilts below $\sim 65^{\circ}$.

\begin{figure}
    \centering
    \includegraphics[width=\linewidth]{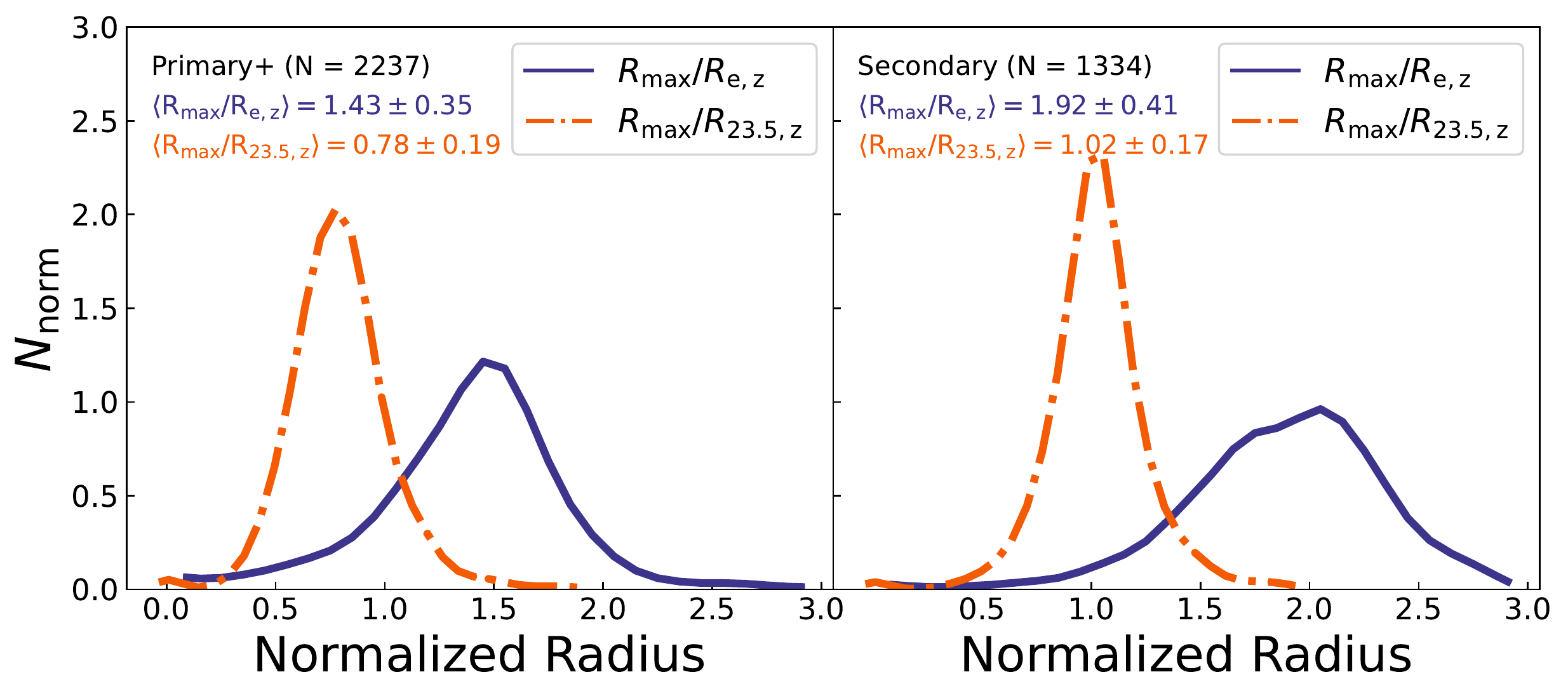}
    \caption{Probability Density Functions for the MaNGA RC extents in terms of various normalised photometric radii. 
    $R_{\rm max}$, $R_{\rm e, z}$, and $R_{\rm 23.5, z}$ are, respectively, the luminosity-weighted maximum radius for the MaNGA IFU, the effective radius, and the isophotal radius measured at a surface brightness level of 23.5 mag\,arcsec$^{-2}$, all reported in the \textit{z}-band. 
    $R_{\rm e, z}$, and $R_{\rm 23.5, z}$ were obtained from DESI photometry of MaNGA galaxies (\protect\photopaper). 
    The left and right panels show primary (+ colour enhanced) and secondary samples, separately. 
    The insets give the median RC extents with an error of 1$\sigma$.}
    \label{fig:rc_extent}
\end{figure}

The coupling of the MaNGA spectroscopic data with our photometry from \photopaper enables us to quantify the physical extent of the MaNGA spectral data.
The MaNGA primary and color-enhanced samples (primary+) have indeed been claimed to probe regions out to 1.5R$_{\rm e}$, and the secondary sample  would extend out to 2.5R$_{\rm e}$\citep{Bundy2015}. 
The effective radii came from the NASA Sloan Atlas (NSA) catalog \citep{Blanton2007}.
The distribution of the extent of the spectroscopic data scaled with various characteristic radii extracted from \photopaper is shown in \Fig{rc_extent}.
The extent of both primary+ and secondary sample is found to be lower than the nominal 1.5 and 2.5R$_{\rm e}$.
This is because our non-parametric approach and more sensitive DESI imaging yield $\sim$0.3\,mag more light for MaNGA-LTGs (\photopaper) than conventional SDSS imaging.
Indeed, \photopaper's effective radii are larger than those from the NSA catalog.
As a result, the MaNGA primary+ and secondary samples extend out to 1.43 $\pm$ 0.35\,R$_{\rm e}$ and 1.92 $\pm$ 0.41\,R$_{\rm e}$, respectively. 

\Fig{rc_extent} also shows the MaNGA RC extents scaled by isophotal radius.
The isophotal (or mass density) measurements for galaxy sizes are more robust, accurate, and reproducible.
Furthermore, the use of isophotal radii (either calibrated by surface brightness and/or stellar mass surface density) to measure structural parameters reduces the scatter of 
related structural scaling relations (\citealt{Hall2012, Trujillo2020}, \photopaper).
In term of isophotal size, measured at the 23.5\,\magss isophote, the MaNGA spectroscopic data extends out to 0.78 $\pm$ 0.19 (1.02 $\pm$ 0.17) R$_{23.5}$ for the primary+ (secondary) samples, respectively.

\subsubsection{Error estimation}

We use ``Jackknife resampling'' \citep{jackknife} to estimate the fit parameter errors of our dynamical models (\Eq{vel_model}). 
A spaxel is randomly selected and all spaxels within 2.5\arcsec radius are removed to produce a new truncated velocity map.
The new velocity map is then fit again with \Eq{vel_model} to calculate truncated fit parameters.
The Jackknife resampling process is repeated 100 times. 
The error on the fit parameters is the 16-84 per\,centile range for the 100 Jackknife runs. 
These errors are also reported in the (public) catalogue of dynamical properties presented in \app{cat}. 
No velocity data were rejected during the fitting process; rather, galaxies were excised from the sample (if at all) during the error estimation process.

\subsubsection{Quality cuts}

Our MaNGA LTG sample \citep{mldl} consists of 5166 galaxies for which DESI-\textit{grz} imaging and robust photometry through \ap is available .
To obtain RCs for MaNGA LTGs, we only use galaxies which achieve photometric inclinations between 30-80 degrees. 
This inclination cut ensures robust inclination corrections for the RC.
For galaxies with DESI photometric data and favourable inclinations, we run our velocity model to extract RCs.
For the galaxies with both photometry and RCs, we only use systems where the seven fits parameters for the velocity models have errors $<20\%$.
In the end, we are left for the remainder of this study with a final set of 2368 galaxies which have robust photometry and RCs.

\subsection{Parameter Extraction and Correction} \label{sec:parameter}

While the photometric (\photopaper) and dynamical catalogues presented here are highly versatile, we are primarily concerned with the sizes, stellar mass, velocity metrics, dynamical masses, and stellar mass surface density within a physical radius (e.g., 1\,kpc).
 
The photometric properties were extracted and corrected using methods described in \photopaper.
These are briefly described below; see \photopaper for more details.

\subsubsection{Corrections}
All SB profiles used in this study were corrected for Galactic and internal extinction, as well as for cosmological K-corrections.
The Galactic extinction corrections were retrieved from the NSA catalogue \citep{Blanton2011}, originally taken from \cite{Schlegel1998}. 
These corrections are available for the optical \textit{grz} photometric bands.
The cosmological K-corrections were calculated using templates from \cite{Blanton2007}.

Our corrections for internal dust extinction and geometry/inclination were applied after extracting the structural parameters. 
The inclination corrections were calculated using a linear fit between each structural parameter and the log of the cosine of the inclination \citep{Giovanelli1994}.
The inclination correction can be written as: 
\begin{equation}
    \log X_{0}=\log X_{i}+\gamma\log(\cos i),
\end{equation}
\noindent where $X_{0}$ is the galaxy property corrected to face-on view, $X_{i}$ is the observed galaxy property, $\gamma$ is the inclination correction factor, and $i$ is the observed inclination of the galaxy disk on the sky.
The photometric inclination, $i$, is related to the projected axis ratio ($b/a$) of the object as
\begin{equation}
    \cos^2i = \frac{(b/a)^2-q_0^2}{1-q_0^2},\qquad (q_0 = 0.13)
    \label{eq:incl}
\end{equation}
where $q_0$ is the assumed stellar disk thickness \citep{Hall2012}.
For more details about the various corrections to the MaNGA photometric data, see \photopaper.

The RCs are also corrected for inclination by dividing the sin of the disk inclination (see \Eq{vel_model}). 
It should be noted that the kinematic inclinations are independent of the photometry as the former is a fit parameter in our dynamical model. 
The difference between kinematic and photometric inclinations was found to be $\sim 5\pm 9^{\circ}$; where the error presents the standard deviation.

\subsubsection{Parameter extraction}

The sizes, stellar masses, and surface densities were estimated using the photometric data.
For greater compatibility with simulations (which predict mass rather than light), we convert all of our light-weighted properties into mass-weighted properties.
Stellar mass estimates for the MaNGA galaxies were calculated using various stellar mass-to-light color relations \citep[MLCRs][]{Courteau2014}.
To achieve the maximum robustness for our stellar mass estimates, five different MLCRs were averaged from \cite{RC15}, \cite{Z17}, and \cite{Benito2019}.
The stellar mass estimates use $g-r$ and $g-z$ optical colours and luminosities from all three DESI bands. 
As a result, the stellar masses are calculated as the average of 30 different measurements.
The average scatter on stellar mass measurements from different MLCRs is $\sim 0.2\,{\rm dex}$.
This is consistent with systematic errors in stellar mass based on extensive MLCRs calibrations reported in \citet{Courteau2014} and \citet{Roediger2015}.

The physically-motivated galaxy sizes used in this study represent the semi-major axis of various isophotal levels, measured in mag\,arcsec$^{-2}$ or M$_{\odot}$\,pc$^{-2}$.
As shown by \cite{Trujillo2020}, such a representation of sizes yields a tighter scatter for the size--mass relation. 

We use the quantity, $\Sigma_1 = M_{\rm *, 1\,kpc}/\pi$ [$M_{\odot}$\,pc$^{-2}$], as a representation of stellar mass surface density within a projected radius; specifically 1\,kpc.

Velocity metrics were calculated by interpolating/extrapolating the inclination-corrected RCs at various radii.
Finally dynamical masses, $M_{\rm dyn}(R)$, are calculated using the velocity, $V(R)$, at projected radius, R, assuming a spherical halo:
\begin{equation}
    M_{\rm dyn}(R) = \sqrt{\frac{V^2(R)R}{G}}.
    \label{eq:m_dyn}
\end{equation}

The error in the dynamical mass at radius R is calculated in quadrature as 
\begin{equation}
    \delta M_{\rm dyn}(R) = \Bigg[ \Bigg(\frac{2V(R)R\delta V(R)}{G}\Bigg)^2 + \Bigg( \frac{V(R)^2\delta R}{G}\Bigg)^2\Bigg]^{1/2}.
    \label{eq:m_dyn_err}
\end{equation}
In \Eq{m_dyn_err}, $\delta V(R)$ and $\delta R$ are the measured uncertainties.

\subsection{Structural Scaling Relations} 
\label{sec:compare}

\begin{figure}
    \centering
    \includegraphics[width=\linewidth]{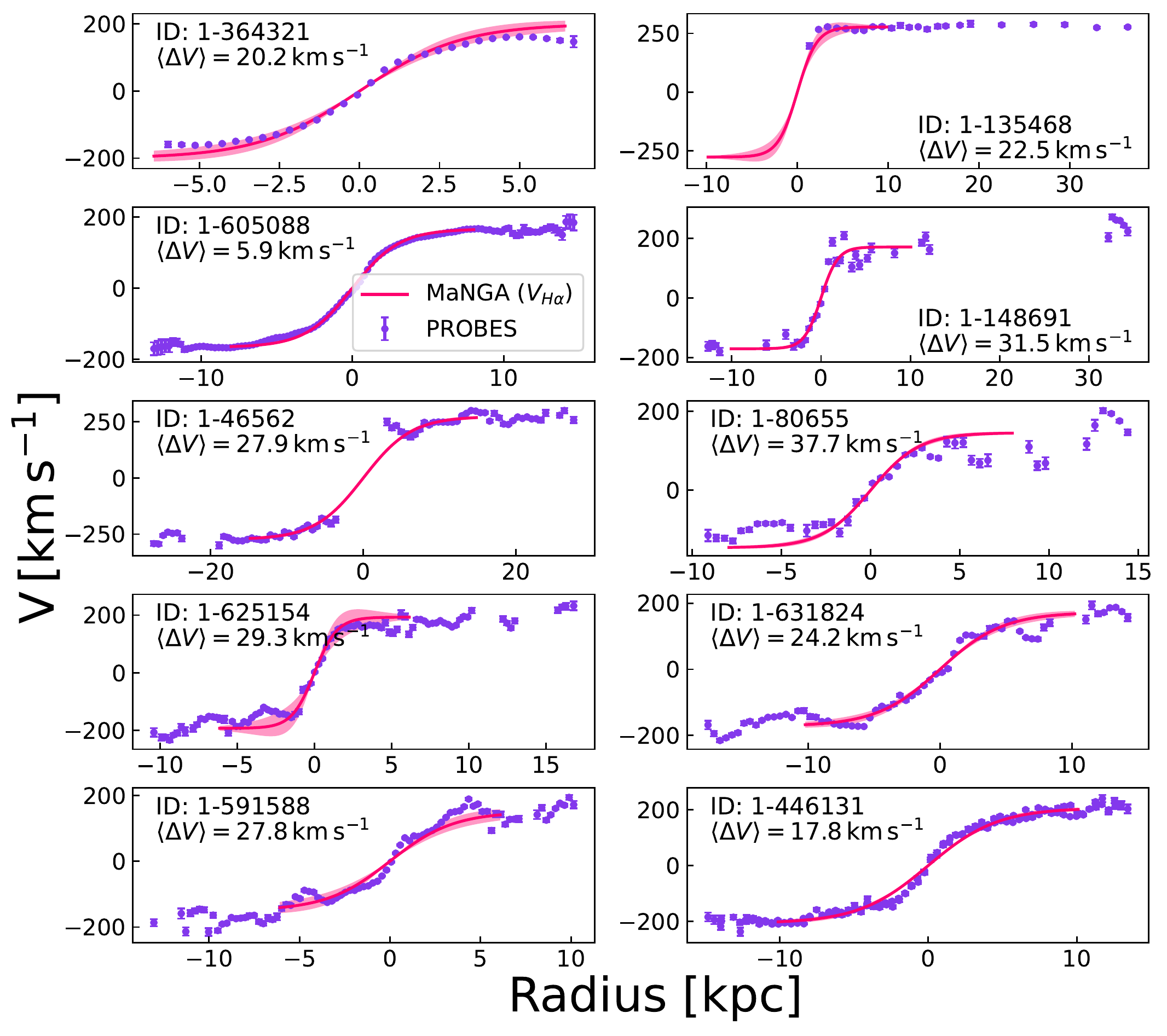}
    \caption{Comparison of inclination-corrected RCs for the ten galaxies in common between MaNGA and the PROBES sample \citep{Stone2020}.
    Each panel compares the long slit \ha RC (purple points with error bars) from PROBES with the RC (pink solid lines with shaded regions) extracted from the MaNGA \ha velocity map. 
    The inset text gives the MaNGA object ID and the RMS between the MaNGA/PROBES data sets.}
    \label{fig:probes_manga}
\end{figure}

We confirm the validity of our MaNGA RCs by comparing them with similar independent observations for the same object.
\Fig{probes_manga} shows such a comparison for the ten galaxies overlapping between MaNGA and the PROBES (``Photometry and Rotation Curve Observations from Extragalactic Surveys'') sample \citep[hereafter PROBES;][]{Stone2019, Stone2020, Stone2022}.
MaNGA's spectroscopic data were taken with an IFU, while most PROBES RCs were obtained via \ha long-slit spectra (except 1-135468 taken in \hi).
While the MaNGA RCs can be extrapolated to infinity (as a result of the tanh fit), the extent of the models shown in \Fig{probes_manga} is limited to the IFU spatial coverage.
The inclinations for the PROBES-MaNGA comparisons use the same photometric ellipticities obtained via AutoProf (\photopaper). 
This ensures a uniform (``apples-to-apples'') comparison for the RCs.
The pair of 10 RCs between PROBES and MaNGA trace each other quite well; on average the RCs from both samples differ by ${\sim}24$\,km\,s$^{-1}$ or ${\sim}0.1\,V_{\rm max}$.
The match is nearly perfect for galaxies (MaNGA-ID) 1-446131 and 1-605088. 
Galaxy (MaNGA-ID) 1-135468 shows a great match between the MaNGA IFU and \hi observations (which naturally extends further). 
While PROBES RCs have typically greater spatial extent, MaNGA velocity maps provide better coverage in the central parts of disk galaxies and take advantage of the two dimensional velocity fields. 

At this point, it is worth mentioning that our velocity models are not corrected for any beam-smearing correction.
Our decision is two-fold. Firstly, the excellent agreement between the MaNGA and PROBES rotation curves presented in \Fig{probes_manga} confirms that the effect of beam smearing is minimal for our galaxy sample. 
Secondly, for the spatially resolved analysis, we ignore the inner two most data points (corresponding roughly to 2 arcseconds). 
This lessens any beam smearing effect in the centrals parts of galaxies. 
Beam smearing ought to be more pronounced for stellar velocity dispersion measurements, an effect that we shall revisit elsewhere.

\begin{figure}
    \centering
    \includegraphics[width=\linewidth]{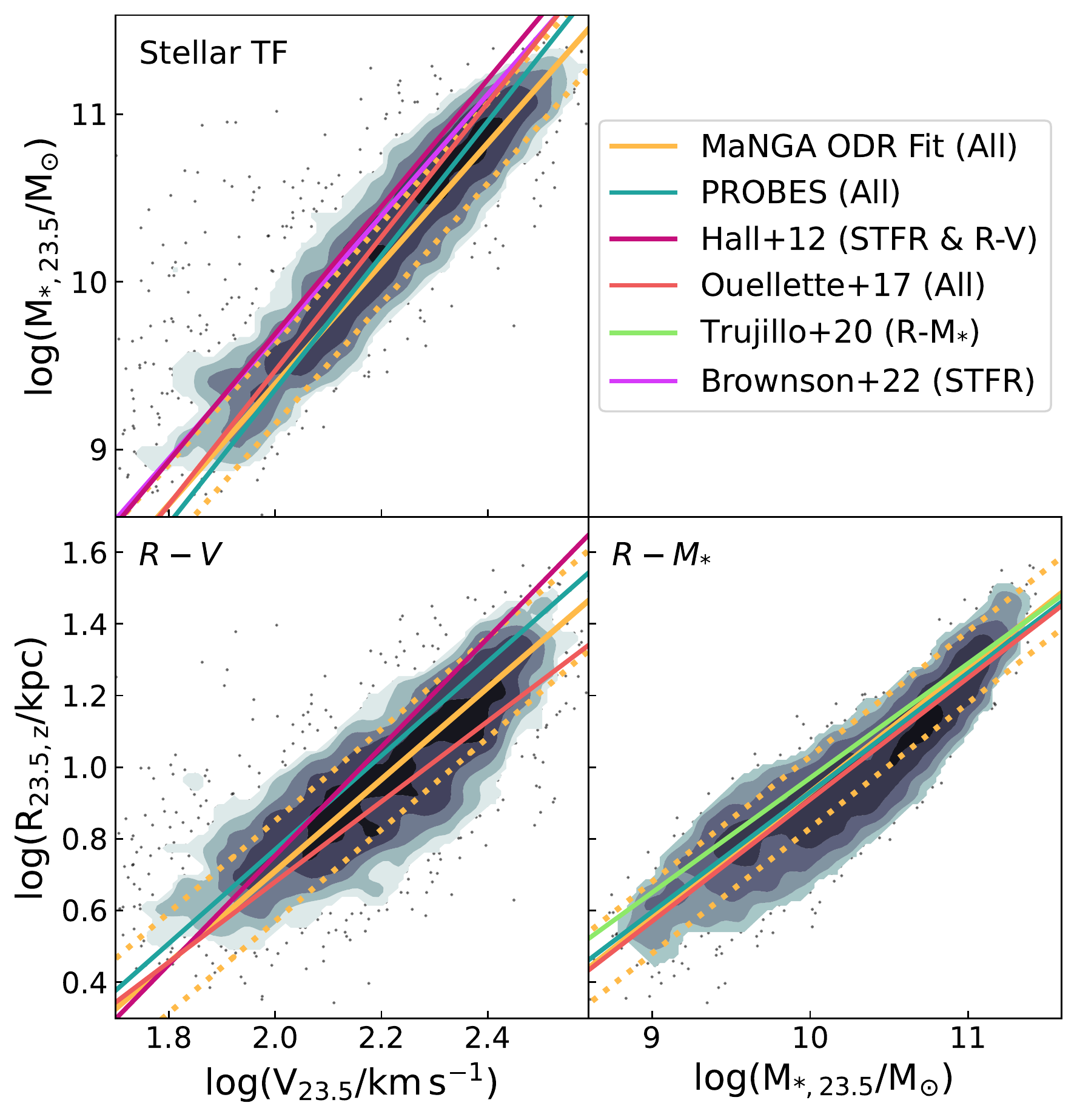}
    \caption{MaNGA VRM$_*$ scaling relations measured at an isophote of 23.5\,mag\,arcsec$^{-2}$ in the $z$-band. 
    The three panels show the stellar TFR (top-left), R--V relation (bottom-left),  
    and the R--M$_{*}$ relation (bottom-right). 
    In each panel, the density contours show the underlying MaNGA data fitted with a linear orthogonal distance regression (orange solid line; the 1$\sigma$ scatter is shown as an orange dotted line); 
    the points are presented in the low density regions. 
    Each scaling relation is compared with various literature sources.
    The legend (top-right) highlights those sources with relevant scaling relations in parenthesis.}
    \label{fig:manga_vrm}
\end{figure}

\begin{table*}
\begin{center}
\begin{tabular}{@{}lccccccc@{}}
\toprule
Scaling Relation             & N & $y$                    & $x$                    & $\alpha$      & $\beta$         & $\sigma_{\rm obs}$ & $\sigma_{\perp}$ \\ 
(1)                          & (2)& (3)                    & (4)                    & (5)           & (6)            & (7)                & (8)              \\ \midrule
Stellar Tully-Fisher         & 2368& $\log M_{\rm *, 23.5}$        & $\log V_{23.5}$ & $3.59\pm0.04$ & \phantom{-}$2.21\pm0.08$  & $0.24\pm0.01$      & 0.06$\pm0.01$\\
Projected size--velocity               & 2368& $\log V_{23.5}$     & $\log R_{23.5, z}$        & $1.28\pm0.02$ & $-1.85\pm0.05$ & $0.14\pm0.01$      & 0.09$\pm0.01$\\
Projected size--stellar mass & 2368& $\log R_{\rm 23.5, z}$ & $\log M_{\rm *, 23.5}$     & $0.35\pm0.01$ & $-2.57\pm0.03$ & $0.10\pm0.01$      & 0.09$\pm0.01$\\
\bottomrule
\end{tabular}
\end{center}
\caption{VRM$_*$ scaling relations for the 2368 MaNGA galaxies. 
Each scaling relation is modelled as $y=\alpha x+\beta$ through orthogonal distance regression. 
Column (1) gives the scaling relation, column (2) shows the number of points fit; columns (3-4) present the $x$ and $y$ variables for the fit; columns (5-6) show the slope ($\alpha$), zero-point ($\beta$), and the observed scatter ($\sigma_{\rm obs}$), respectively; and, column (8) gives the orthogonal scatter ($\sigma_{\perp}$) for the scaling relation.}
\label{tab:manga_fits}
\end{table*}

The full catalogue enables us construct scaling relations which can also be compared with the literature. 
\Fig{manga_vrm} presents the VRM$_*$ scaling relations for $\sim$2300 LTGs compared with other sources.
The orthogonal distance regression (hereafter ODR) linear fits and scatters for each relation are given in \tab{manga_fits}.
In general, our photometry and RC modelling yield scaling relations that match other studies. 
For the sake of uniformity and unless otherwise stated, the selected comparisons in \Fig{manga_vrm} use isophotal radii measured at the 23.5 mag\,arcsec$^{-2}$ isophote.
In what follows, we briefly discuss and compare each scaling relation.

\subsubsection{Projected size -- stellar mass relation}

The MaNGA size-mass (R -- M$_{*}$) relation, shown in the bottom-right panel of \Fig{manga_vrm}, has an ODR slope of $\log (R_{23.5, z}/{\rm kpc})=(0.35\pm0.01)\log (M_{*,23.5}/{\rm M_{\odot}})+(-2.57\pm0.03)$ and an observed scatter of $0.10\pm0.01$ dex.
That relation, the same as that presented in \photopaper, can be compared with similar ones presented by \cite{Ouellette2017}, \cite{Trujillo2020}, and \cite{Stone2020} for different samples. 
For instance, \cite{Ouellette2017} used SDSS \textit{i}-band photometry for the SHIVir survey of Virgo cluster galaxies to derive a R -- M$_{*}$ relation with a similar ODR slope of $0.34\pm0.02$, shown as the dark orange solid line, though with a larger scatter of 0.15~dex. 
That larger scatter is due to a much lower sample size (69 galaxies) in the SHIVir survey. 
\cite{Stone2020} also constructed a R -- M$_{*}$ relation for 1152 PROBES galaxies with structural parameters measured using similar methods as ours. 
Both analyses of the PROBES and MaNGA data sets used DESI imaging and SB profiles extracted through the \textsc{Autoprof}.
It is therefore not surprising that their ODR slope of $0.334^{0.009}_{0.004}$ and scatter of $0.099^{0.002}_{0.003}$ should be so close to ours for MaNGA. 
Any (very small) difference ought to be explained by the use of two different samples. 

Studying 464 LTGs from the SDSS IAC Stripe82 survey, \cite{Trujillo2020} found an isophotal ODR slope of $0.318\pm0.014$ (light green solid line) and a scatter of $0.087\pm0.005$.
Isophotal sizes are sensitive to the choice of photometric bands; while we use \textit{z}-band, \cite{Trujillo2020} used the SDSS-\textit{i} band.
The difference in the fit parameters are a result of choice of photometric band for our respective analysis.

\subsubsection{Projected size -- velocity relation}

The size -- velocity (RV) relation for MaNGA LTGs, shown in the bottom-left panel in \Fig{manga_vrm}, has an ODR fit of $\log (R_{23.5, z}/{\rm kpc})=(1.28\pm0.02)\log (V_{23.5}/{\rm km\,s^{-1}})  + (-1.85\pm0.05)$ with an observed scatter of $0.14\pm0.01$.
The PROBES galaxies also benefit from long-slit \ha RCs and yield an RV relation whose parameters are closely matched with MaNGA's; $\alpha=1.325^{0.074}_{0.034}$ and $\sigma_{\rm obs}=0.128\pm0.004$.
The RV relation for the SHIVir galaxies \citep{Ouellette2017} has a much smaller slope of $1.1\pm0.1$. 
The difference could be due to environmental effects, or small number statistics. 
The latter actually thwarts a conclusive answer. 
\cite{Hall2012} used SDSS photometry and rotational velocities from integrated \hi linewidths from \cite{Springob2005, Springob2007} to find a steeper RV slope of $1.518\pm0.065$ and an observed scatter of 0.152~dex. 
While our scatters agree within the error, the slopes for our respective RV relations differ significantly. 
While integrated \hi linewidths are often shown to reproduce various well-defined spatially-resolved rotational velocities \citep{Courteau1997}, an exhaustive study should be conducted to characterize the spatial coverage of the \hi linewidths used by \cite{Hall2012}. 
The latter is beyond the scope of this paper.

Different velocity measurements can definitely result in different slopes for our respective R -- V relation.
Other sources of uncertainties include distance measurements. 

\subsubsection{Stellar Tully -- Fisher relation}

The combination of our photometry (\photopaper) and RC analysis yields a stellar Tully-Fisher (STFR), or M$_*$--V relation,
which informs us about the cross-talk between the stellar and dark matter in MaNGA galaxies \citep{Dutton2010, Hall2012, Ouellette2017, Stone2020, Ortiz2020}.
The STFR for MaNGA LTGs, measured at R$_{23.5}$, shown in the top-left panel of \Fig{manga_vrm}, is described by $\log (M_{*,23.5}/{\rm M_{\odot}}) = (3.59\pm 0.04)\log (V_{23.5}/{\rm km\,s^{-1}})+(2.21\pm 0.08)$ with an observed scatter of $0.24\pm0.01$ dex.
This ODR fit slope agrees with theoretical derivations of $\alpha \sim 3-4$  \citep{Courteau1997}.
Moreover, of the three VRM scaling relations studied here, the STFR is the tightest with a small orthogonal scatter ($\sigma_{\perp}=\sigma_{\rm obs}/(1+m^{2})^{0.5}$), as shown in \Tab{manga_fits}. 
The study of 1152 PROBES galaxies by \cite{Stone2020} yielded a STFR slope of $3.997^{0.144}_{0.039}$ and an observed scatter of $0.248^{0.009}_{0.007}$ dex.
Similarly, \cite{Ouellette2017} found a STFR slope of $3.99\pm0.18$ for SHIVir galaxies.
These two STFRs are actually steeper than the one found here for MaNGA galaxies. 
This discrepancy can be explained by the sample differences; SHIVir and PROBES sample consists of more dwarf galaxies than MaNGA.
As is seen in the Baryonic Tully-Fisher relation, dwarfs have a different stellar TFR slope until corrected for gas mass.

\cite{Brownson2022} also studied the correlation between dynamical tracers and star formation quenching in galaxies with MaNGA galaxies. 
Their modelling of stellar velocities used a formalism similar to ours with \Eq{vel_model}, \cite{Brownson2022} yiedling a STFR slope of 3.62$\pm$0.13.
The stellar velocity used by \cite{Brownson2022} would further suffer from asymmetric drift and would also be a source of differences between our respective STFR.
The latter is certainly a closer match to our \ha velocity-based STFR slope than those measured by \cite{Ouellette2017} or \cite{Stone2020}.
However, the scatters differ, 0.24 dex in this study versus 0.09 dex in \cite{Brownson2022}, as a result of the definition of scatter interquartile range versus root mean square error.
The zero-points also differ by $\sim$0.2 dex, due to the different stellar measurements and various assumptions such as the choice of IMFs and the location of stellar mass measurements.

\subsection{Selection Functions}
A discussion about the selection functions of the observed and simulated galaxy sample is warranted. 
For the sake of uniformity within our data-model comparison, we down-sample the complete observed MaNGA sample to a morphologically-limited massive LTG sample \citep{mldl}. 
While the conclusions of the comparison presented in this paper remain valid, they are only limited to the structure of massive disk systems. 
Our aim is not to achieve a volume-complete or bias-corrected observed sample for comparison with simulations but rather to examine the ability of current simulations to reproduce various observed properties and scaling relations for a segment of the galaxy population.
Similar data-model comparisons have indeed already served as a guide for cosmological simulations \citep{Starkenburg2019, Nanni2022, Cannarozzo2022, Frankel2022, Goddy2023}. 

\section{NIHAO Galaxy Formation Simulations} 
\label{sec:nihao}

\begin{figure*}
    \centering
    \includegraphics[width=\columnwidth]{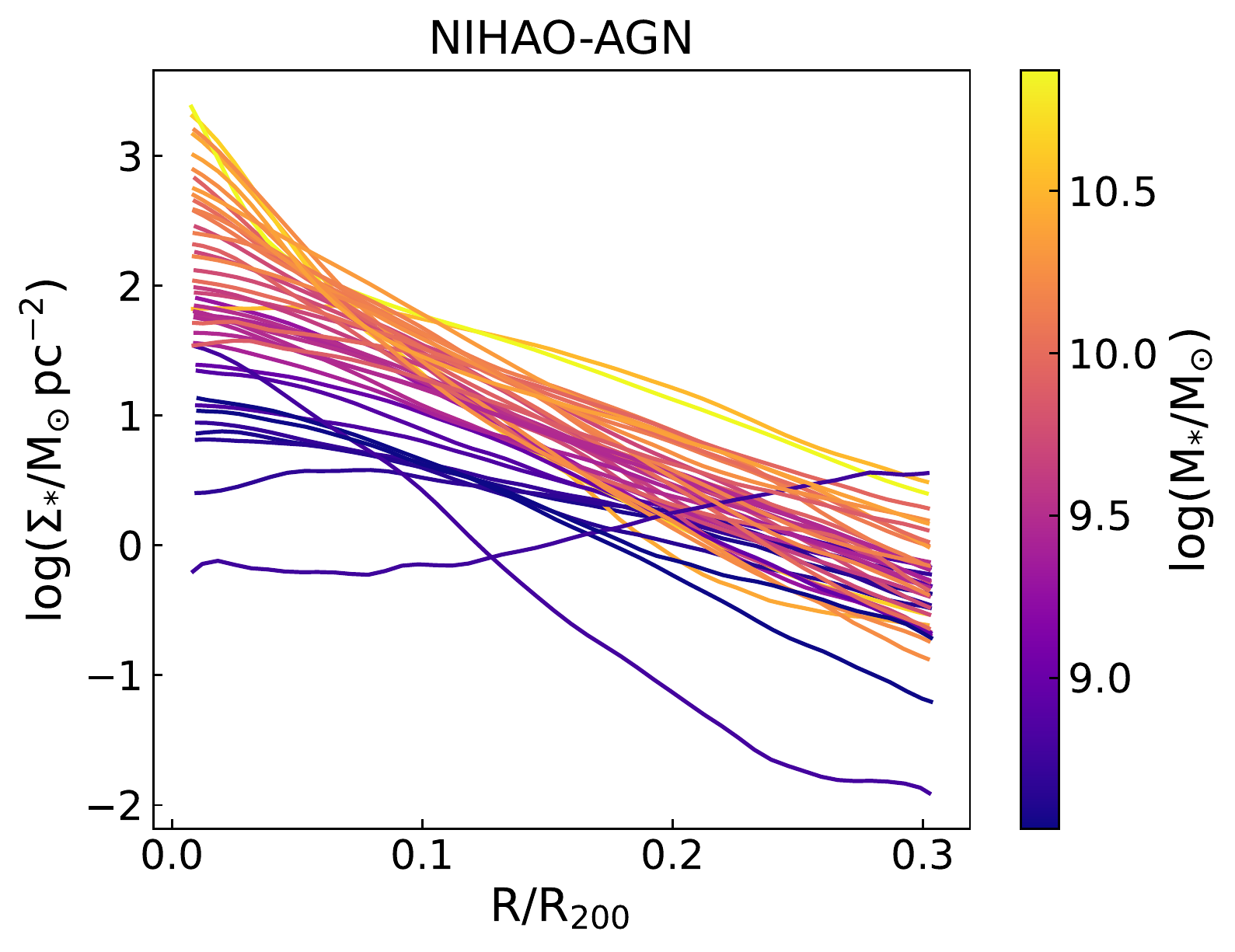}
    \includegraphics[width=\columnwidth]{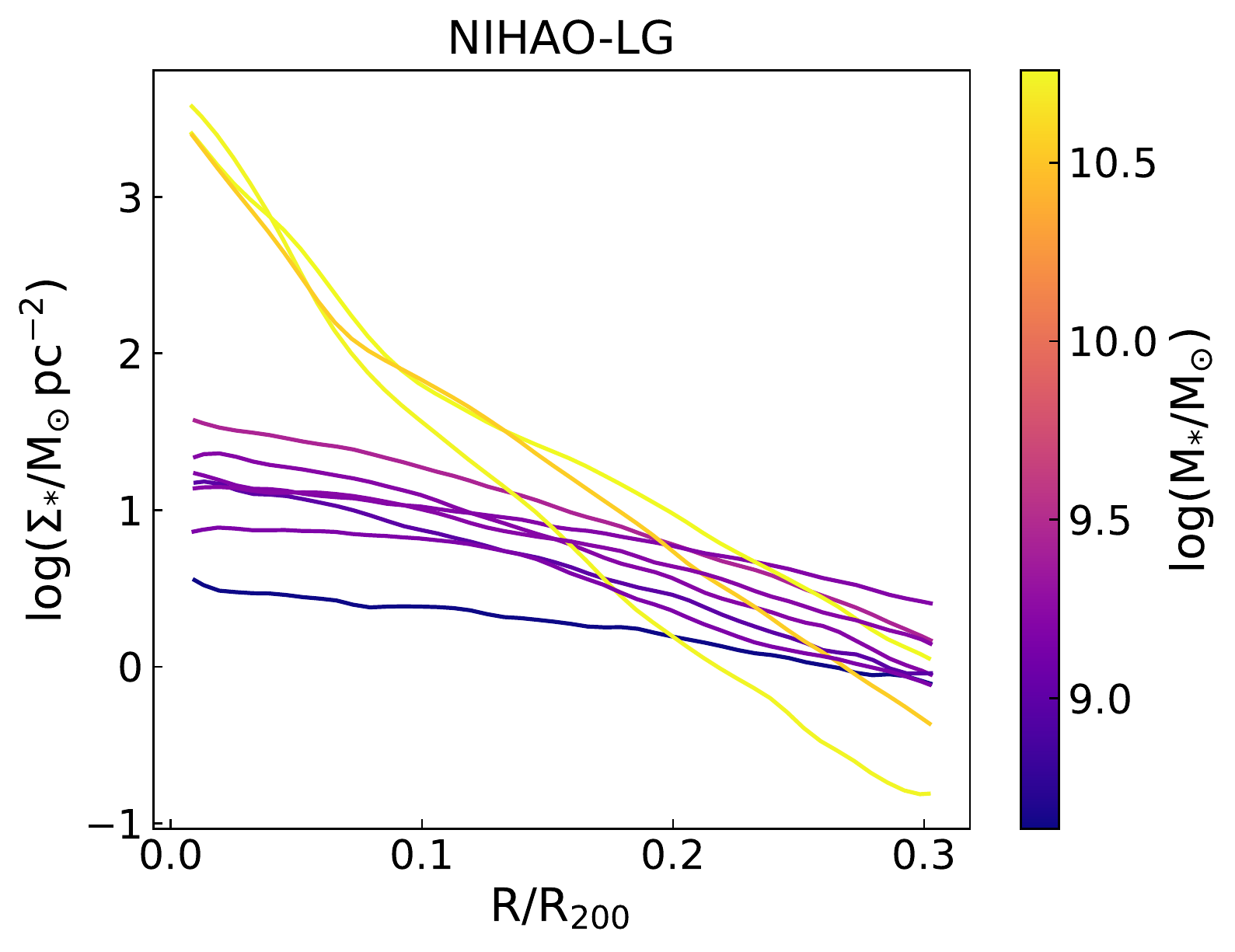}
    \caption{Stellar mass surface density profiles for NIHAO-AGN (left panel) and NIHAO-LG (right panel) galaxies. 
    The colour of each line represents  the stellar mass of each galaxy at $z=0$. 
    The radial axis is normalized by $R_{\rm 200}\,(= 600H_{0}^2/8\pi G)$.
    The stellar mass density profiles are calculated for the halo aligned face-on.}
    \label{fig:nihao_sig_prof}
\end{figure*}

\begin{figure*}
    \centering
    \includegraphics[width=\columnwidth]{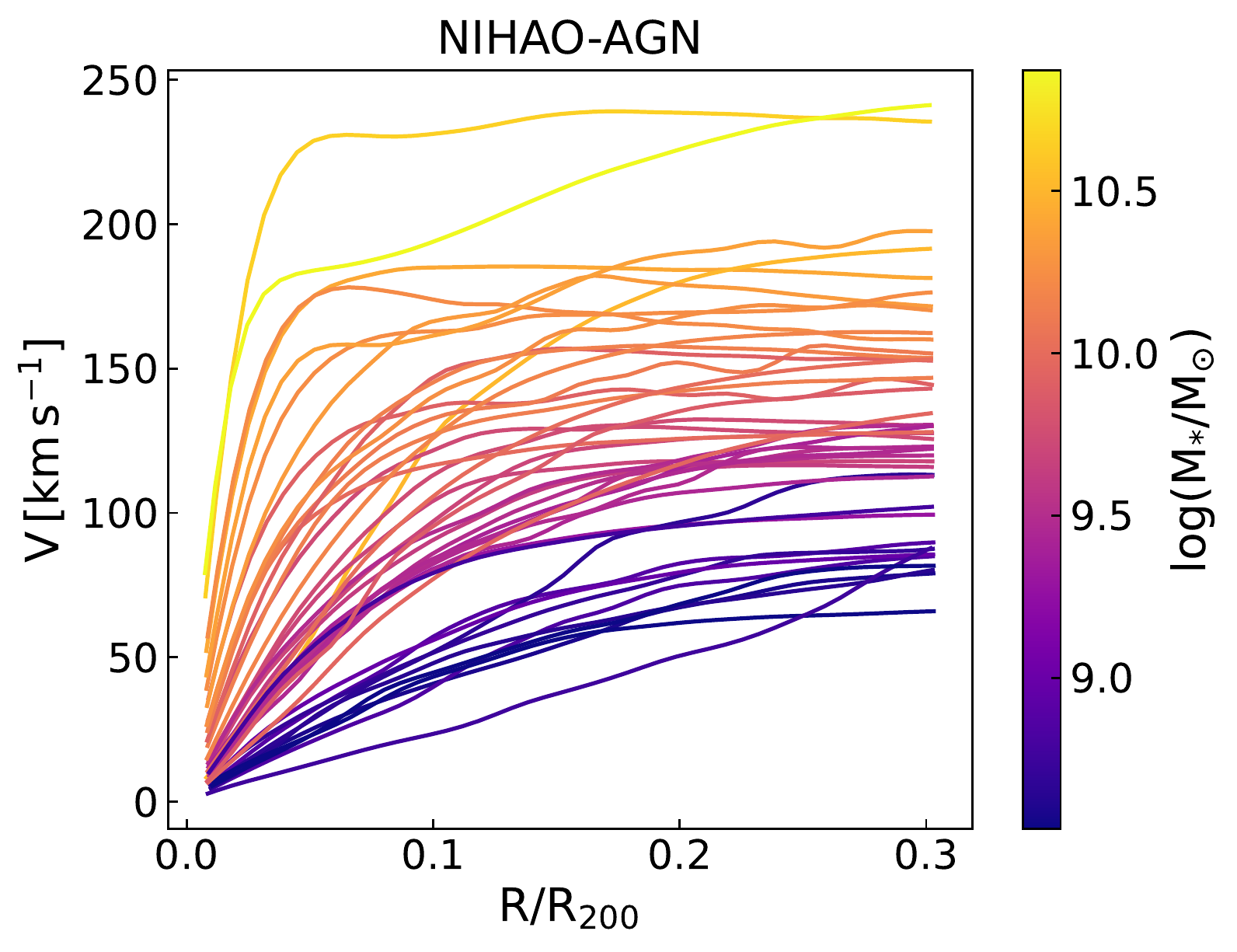}
    \includegraphics[width=\columnwidth]{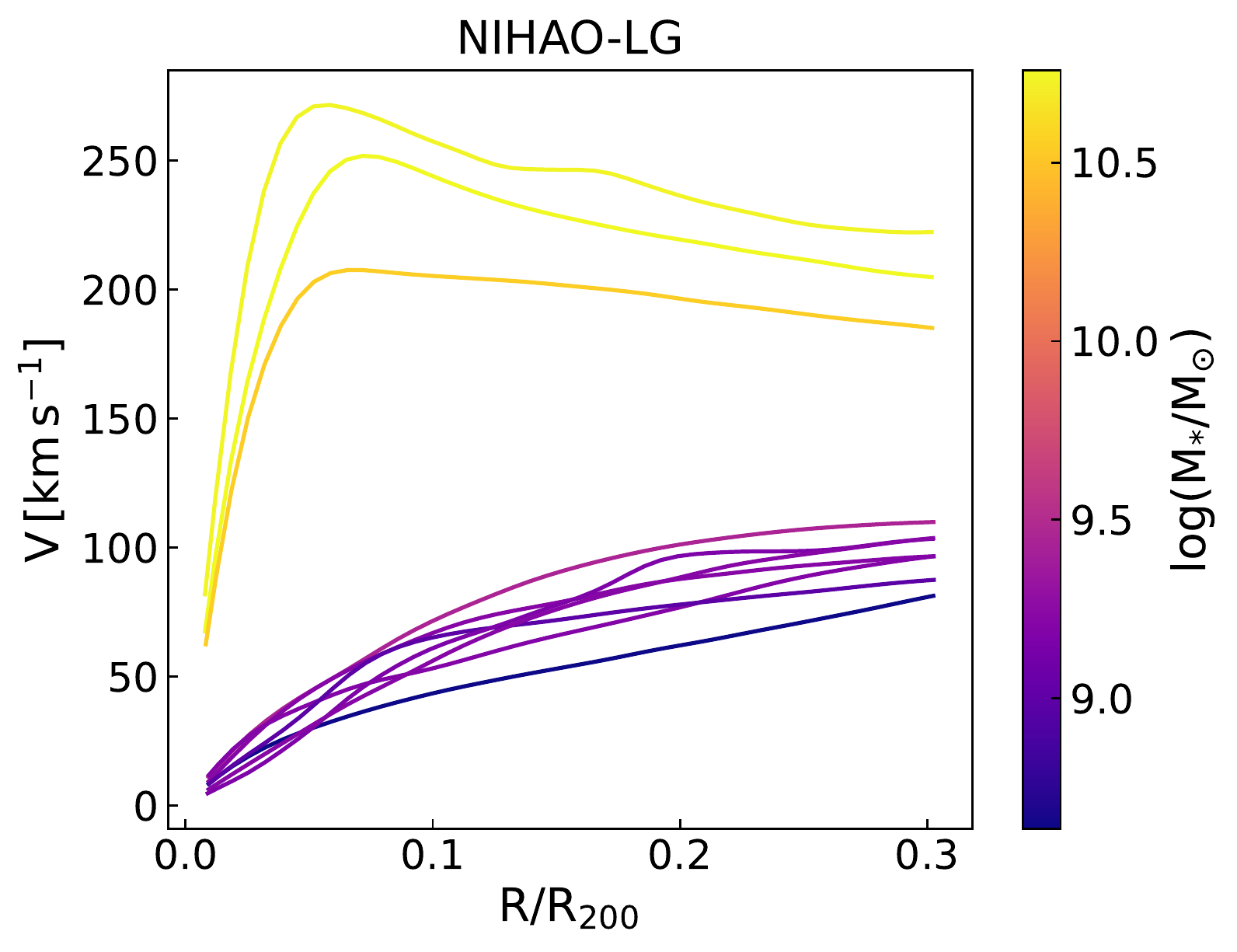}
    \caption{RCs for NIHAO-AGN (left panel) and NIHAO-LG (right panel) galaxies. 
    The colour of each line represents the stellar mass of each galaxy. 
    The radial axis is normalized by $R_{\rm 200}$.
    RCs are calculated for halos viewed edge-on.}
    \label{fig:nihao_rc}
\end{figure*}

Having established a solid baseline of observed disk galaxy scaling relations for MaNGA sample, we are now poised to make direct comparisons with numerical simulations of galaxy formation in order to tease out any differences that could either indicate limitations of our theoretical formalism or biases in the way that structural parameters are being measured.
For the sake of this comparison,  we use galaxies from the ``Numerical Investigation of a Hundred Astrophysical Objects" (NIHAO) cosmological zoom-in simulations \citep{nihao_main}.  
The NIHAO simulations were run with a flat $\rm \Lambda$CDM cosmology with parameters from the \textit{Planck Satellite Investigation}: $\rm H_0 = 100h\,km\,s^{-1}\,Mpc^{-1}$ with $\rm h =0.671,\, \Omega_m=0.3175,\, \Omega_{\Lambda}=0.6824,\, \Omega_b=0.049,\, \sigma_8=0.8344$ and $\rm n=0.9624$ \citep{planck14}.
The hydrodynamics used an updated N-body SPH solver {\scriptsize GASOLINE} \citep{Wadsley2017}. 
All NIHAO galaxies were allowed to form stars following the Kennicutt-Schmidt law \citep{Kennicutt1998} with suitable density and temperature thresholds, $\rm T<15000\,K$ and $\rm n_{th}>10.3\, cm^{-3}$. 
Energy is re-injected back into the interstellar medium (ISM) from stars through blast wave supernova feedback. 
Massive stars also ionize the ISM before their supernova explosion referred to as ``early stellar feedback" \citep{Stinson2006, nihao_main}. 
The ``early stellar feedback" (ESF) mode allows 13\% of the total stellar flux to be injected into the ISM. 
This differs from the original prescription of \cite{Stinson2013} to account for increased mixing and conform to the abundance matching presented in \cite{Behroozi2013}. 
For the supernova feedback, massive stars with $\rm 8\,M_{\odot}<M_{star}<40\,M_{\odot}$ inject energy and metals into the the ISM. 
The energy is injected into high density gas and is radiated away due to efficient cooling. 
For gas particles inside the blast radius, cooling is delayed by $30\, Myr$ \citep{Stinson2013}. 

Supermassive black holes (SMBH) and their associated feedback are also included in the latest version of the NIHAO simulation \citep{Blank2019}.
All NIHAO galaxies with halo mass $M_{200}>5\times 10^{10} \rm {M_{\odot}}$ were seeded with a SMBH with mass 10$^{5} \rm{M_{\odot}}$.
SMBH accretion in NIHAO galaxies follows the Bondi-Hoyle-Lyttleton parametrization \citep{Bondi1952} where the accretion is limited by the Eddington rate.
The feedback from SMBH is a result of the accretion where 15 per\,cent of the generated luminosity is deposited into the nearby gas thermal energy.
Simulated NIHAO galaxies have been shown to match various observed galaxy properties and scaling relations \citep{Maccio2016, Obreja2016, Buck2017, Dutton2017}.
However, our detailed analysis below will shed some contrasting light on these assertions.
Some genuine challenges, such as matching the galaxy population diversity, do indeed exist \citep{Frosst2022}.

For the present analysis, we use all NIHAO galaxies with AGN feedback, which are centrals, and that have $\log (M_{*} / \rm{M_{\odot}) > 8.5}$.
The latter stellar mass restriction is made to match the MaNGA stellar mass distribution. 
Out of the 91 galaxies from the NIHAO project, 45 NIHAO objects satisfy the stellar mass criteria.
All NIHAO-AGN galaxies were simulated with a dark matter particle mass $1.38\times10^{7}\,\rm{M_{\odot}}$ with a softening length of 1.86 kpc and gas softening of 782 pc.
To further increase the size of our simulated sample, we use massive galaxies from the NIHAO-LG simulations performed with the same hydrodynamics (no AGN feedback) but constrained to reproduce the Local Group environment  \citep{Arora2022}.
For the NIHAO-LG simulations, we also utilize all galaxies with $\log (M_{*}/\rm{M_{\odot}) > 8.5}$.
NIHAO-LG galaxies were run with a dark matter particle mass of $1.62\times 10^{6}\,\rm{M_{\odot}}$ with a softening length of 860.3\,pc and a gas softening length of 487\,pc.
\Figs{nihao_sig_prof}{nihao_rc} show the individual stellar mass surface density profile and rotation curves for NIHAO-AGN and NIHAO-LG galaxies.
Both NIHAO simulations span a large range of stellar mass surface densities and velocities, similar to the MaNGA observations.

\section{One-Dimensional Comparisons}\label{sec:one_d}

\begin{figure*}
    \centering
    \includegraphics[width=\linewidth]{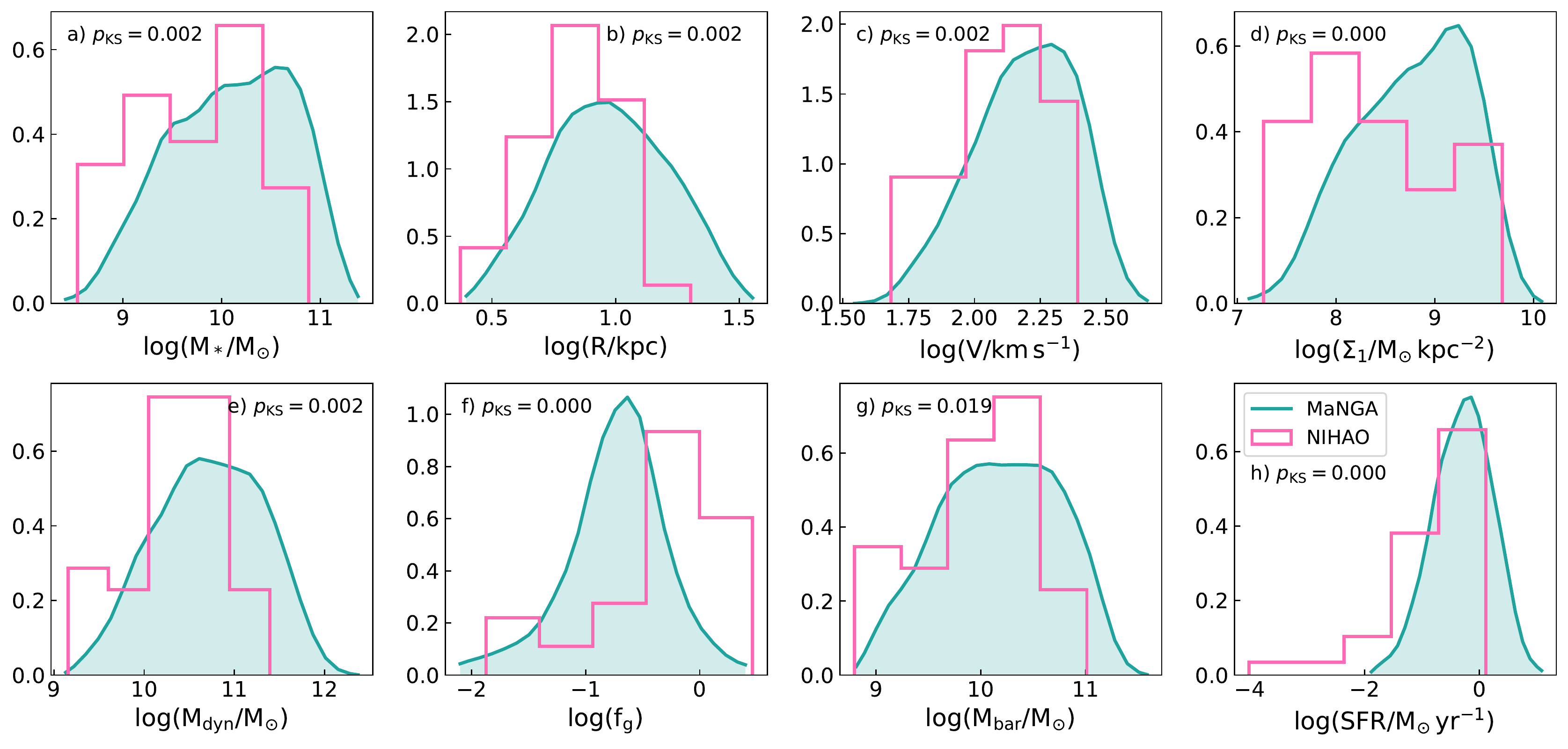}
    \caption{Comparison of the probability density function for eight galaxy properties between MaNGA observations (cyan line) and NIHAO simulations (pink line). The text inset on each panel gives the p-value for the KS test applied to the MaNGA and NIHAO samples. All properties for both observed and simulated galaxies were calculated at a radius corresponding to a stellar surface density of 10\,M$_{\rm \odot}\,$pc$^{-2}$ (with the exception of the gas mass, used for $\log(f_{\rm g})$ and $\log(M_{\rm bar}/{\rm M_{\odot}})$, and SFR obtained by Pipe3D).}
    \label{fig:manga_nihao_hist}
\end{figure*}

With our observed and simulated galaxy samples in place, we can now perform a multi-dimensional comparison.
The simplest data-model comparisons use one dimensional probability density functions (PDF). 
In general, the comparison of data-model PDFs highlights clear areas of agreement/disagreement between galaxy simulations and observations.
\Fig{manga_nihao_hist} presents such a comparison for eight galaxy properties extracted for MaNGA and NIHAO galaxies measured at a radius corresponding to $\Sigma_*=10\,\rm{M_{\odot}\,pc^{-2}}$.
Qualitatively, certain differences between the MaNGA and NIHAO galaxy properties stand out, with the largest differences being found for $M_{*}$, $\Sigma_{\rm 1}$, $f_{\rm g}(\equiv M_{\rm g}/M_{*})$, and star formation rate.
For instance, the NIHAO sample has smaller values of $M_{*}$, $\Sigma_{1}$ and SFR than MaNGA observations. 
NIHAO zoom-in simulations were designed with a roughly uniform stellar mass distribution, which explains the discrepancies for the PDFs in $M_{*}$.
On the other hand the gas fractions, $f_{\rm g}$, in NIHAO have values higher relative to observed galaxies. 

For a quantitative comparison, we perform a Kolmogorov–Smirnov (KS) test on the simulated and observed distributions of galaxy properties. 
The text inset in each panel of \Fig{manga_nihao_hist} shows the p-value for the KS test applied to the MaNGA and NIHAO samples. 
Generally, a value of $p_{\rm KS}<0.05$ is indicative of two samples being drawn from statistically different samples.
This is indeed what we find for all eight galaxy properties shown in \Fig{manga_nihao_hist}.
However, the disagreement between observations and simulations is largely due to the incompleteness of our samples.
The NIHAO and MaNGA galaxy samples were designed to have a roughly uniform stellar mass distribution with a non-trivial selection function; both samples are also statistically incomplete in different ways.
While volume corrected statistics exist for the MaNGA sample \citep{Wake2017}, these are not applied here.
Furthermore, most zoom-in simulations involves non-trivial, complex stellar mass distribution functions and large scale observations are often volume incomplete. 
Therefore, caution is advised in the calculation and interpretation of such comparisons.

\section{Galaxy Scaling Relations} \label{sec:sr}

The next level of complexity in data-model comparisons is to use galaxy scaling relations.
Our approach is to construct multiple galaxy scaling relations to build a more complete picture.
This can capture the multi-faceted nature of galaxy formation and avenues of progress for hydrodynamical simulations.
Typically, structural parameters would be measured at a light-based radius such as R$_{23.5,z}$, however isophotal levels are challenging to evaluate for numerical simulations.
To uniformly compare simulated and observed galaxies, we measure all structural properties at a physically-motivated radius defined by a stellar mass surface density. 
Through the DESI photometry of MaNGA galaxies, we find that R$_{23.5,z}$ corresponds to a median $\Sigma_*\sim 13\,\rm{M_{\odot}\,pc^{-2}}$ with 16-84 per\,cent quartile range of 9.5-16.7\,$\rm{M_{\odot}\,pc^{-2}}$.
Therefore, to match closely with previous studies based on R$_{23.5,z}$, we measure all properties at the $\Sigma_*=10\,\rm{M_{\odot}\,pc^{-2}}$ radius.

\subsection{Qualitative Comparisons}

\begin{figure*}
    \centering
    \includegraphics[width=\linewidth]{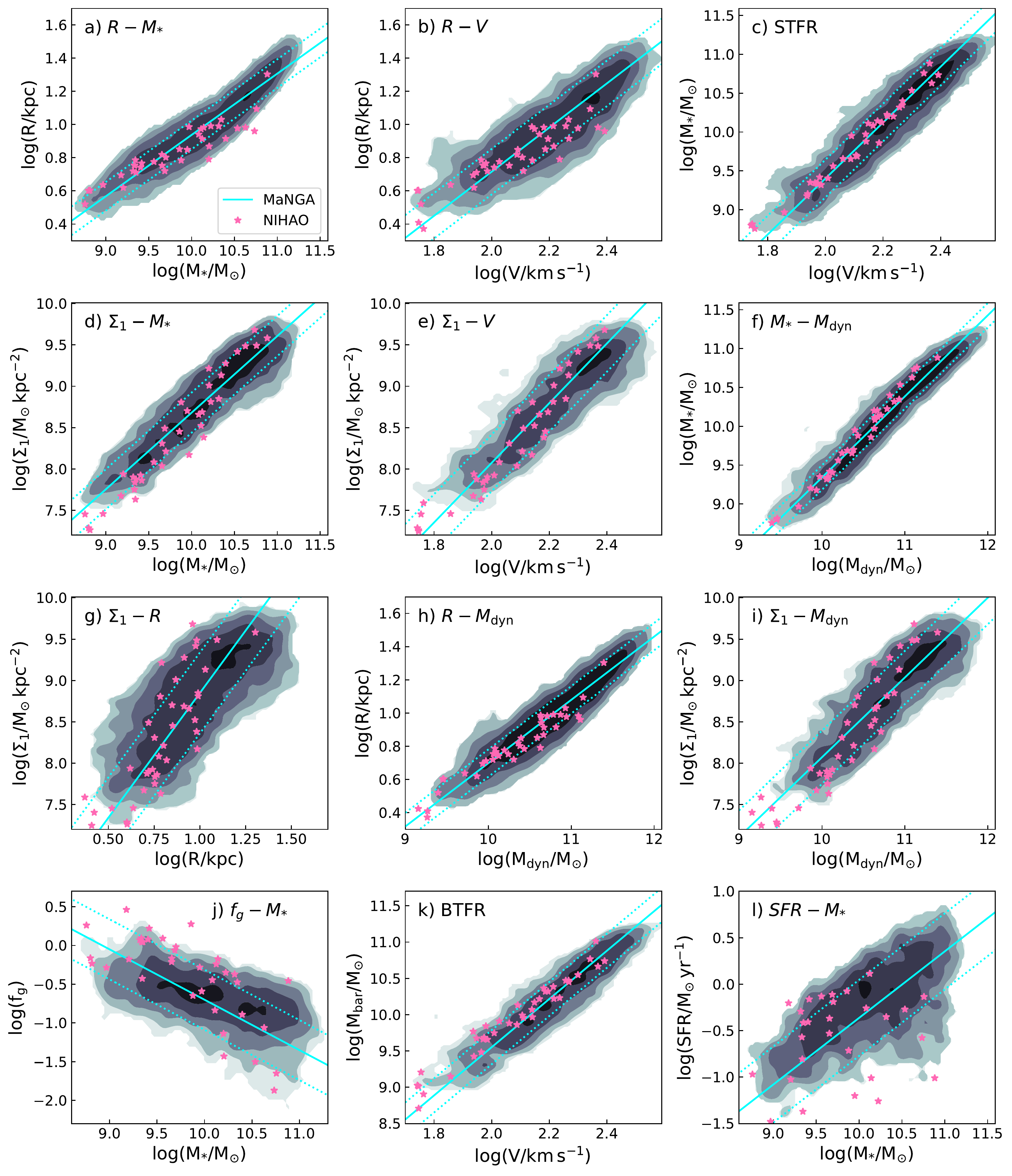}
    \caption{Comparison of 12 scaling relations between MaNGA and simulated NIHAO galaxies. The density contours match the underlying MaNGA observations; the cyan solid and dotted lines present the ODR best fit and scatter respectively. 
    The pink stars represent the simulated NIHAO galaxies. 
    All properties for both observed and simulated galaxies are calculated at a radius which corresponds to a stellar surface density of 10\,M$_{\rm \odot}\,$pc$^{-2}$ (with the exception of the gas mass (Panel j) and SFR (Panel l) which are obtained using Pipe3D). 
    The inset label identifies the scaling relation.}
    \label{fig:manga_nihao_compare}
\end{figure*}

\Fig{manga_nihao_compare} presents a qualitative comparison of various structural scaling relations between MaNGA and NIHAO. 
The corresponding fitting parameters are reported in \Tab{manga_nihao_sr} below. 
The top row of \Fig{manga_nihao_compare} shows the structural VRM$_*$ scaling relations; size-stellar mass, size-velocity, and STFR.
Overall, the simulated NIHAO galaxies follow similar relations as the observed galaxies, as the pink stars scatter more or less evenly about the best fit cyan line, but some differences exist.
Most conspicuously, it is seen that simulated disk galaxies, with large sizes, high stellar masses and circular velocities, are underrepresented.
The under-representation of such galaxies is primarily due to NIHAO's selection function.

For the parameter spaces where NIHAO and MaNGA galaxies overlap, both $R - M_{*}$ and $R-V$ relations are in broad agreement between simulations and observations.
With respect to the STFR, BTFR, $M_{\rm *}-M_{\rm dyn}$, and $R-M_{\rm dyn}$, all NIHAO galaxies scatter within observed 1$\sigma$ region.
The comparison of scaling relations involving $\Sigma_1$ also shows that NIHAO galaxies with high stellar mass ($M_{*}\gtrsim10^{10}$) have higher central stellar densities than observed.
All massive NIHAO galaxies consistently lie at or above the upper 1$\sigma$ observed line. 
These high central stellar densities for massive spiral galaxies are indicative of the weaker baryonic feedback which fails to remove material to prevent over-cooling \citep[see also][]{McCarthy2012}.
At the low mass end, simulated NIHAO galaxies have slightly lower $\Sigma_1$ than the observed MaNGA sample (see Panels d, e, g, and i of \Fig{manga_nihao_compare}). 
For low mass galaxies, the strong stellar feedback removes baryons perhaps too efficiently from the central parts to depress $\Sigma_1$ relative to observations. 
The strong stellar feedback in NIHAO also leads to an overall expansion of the halo \citep{Dutton2016} causing slightly larger sizes compared to MaNGA systems for low mass systems.

Finally, the bottom panel of \Fig{manga_nihao_compare} shows comparison of scaling relations for gas masses (Panel j) and star formation rates (SFRs; Panel l). 
Gas masses and SFRs were taken from the Pipe3D outputs for the MaNGA IFU data \citep{Sanchez2016}.
The observed gas masses are estimated from the dust attenuation while taking the oxygen abundance into account, while SFRs are derived from Pipe3D's simple stellar population (SSP) template fitting for the last 100 Myr.
\footnote{For a subset of MaNGA galaxies, neutral hydrogen gas measurements were made through the ALFALFA survey \citep{Haynes2018} and the Robert C. Bryd Green Bank Telescope. 
Details of the MaNGA-HI catalogue are found in \cite{Masters2019} and  \cite{Stark2021}.}
For simulated NIHAO galaxies, gas mass estimates correspond to the total mass of gas particles within a radius where the stellar surface density is 10\,M$_{\odot}$\,pc$^{-2}$.
SFR estimates are an average SFR within the last 100 Myr.
We note that the gas mass estimates between simulations and observations are not uniformly assessed, since Pipe3D uses MaNGA data that are limited to the extent of the IFU (1.5/2.5\,R$_{e}$).
Furthermore, the Pipe3D outputs do not provide errors for the gas mass and SFR estimates, and therefore intrinsic scatter for scaling relations involving these properties cannot be robustly calculated.

Panel j) in \Fig{manga_nihao_compare} shows the fraction of gas ($f_{g}\equiv M_{g}/M_{*}$) as a function of stellar mass for the MaNGA and NIHAO galaxies. 
As expected, low stellar mass galaxies have a larger fraction of gas relative to high stellar mass galaxies \citep{Catinella2010, Cortese2011, Huang2012}. 
In general, and especially for the lower mass systems, the NIHAO simulated galaxies retain more gas than observed MaNGA distributions; the opposite is true at higher masses. 
As stated above, the low and intermediate stellar mass NIHAO galaxies show signatures of excess cooling (\citealt{Arora2022}).
However, the excess gas could also be a result of the mis-match between the observed and simulated gas measurements.
A few massive NIHAO galaxies have reduced gas content, however consistent with MaNGA observations. 

Panel k) of \Fig{manga_nihao_compare} shows the baryonic Tully-Fisher relation (BTFR) for MaNGA and NIHAO galaxies.
Unlike the STFR, which showed a great simulation-observation match (Panel c), the simulated galaxies rotate too slow for a given baryonic mass. 
This is a systematic trend at most masses, but especially at the low mass end, possibly as a result of over-cooling.
The significant discrepancy at the low-mass end of the BTFR has also been observed in other simulation-observation comparisons \citep{Brook2012, McQuinn2022}.

Finally, panel l) features the SFR$-M_{*}$ relation for MaNGA and NIHAO. 
Along with the star formation main sequence \citep[SFMS;][]{Whitaker2012, Cano2016, Hall2018}, some of the MaNGA LTGs are also found in the ``green valley'' \citep{Salim2014}.
As with many other scaling relations presented here, NIHAO galaxies lie on the distribution of observed MaNGA galaxies.
However, for a given mass, some NIHAO galaxies have much lower SFRs than observed.
Again, this is likely cause by the strong stellar feedback within NIHAO galaxies, leading to lower SFRs than observations.
Massive NIHAO galaxies with high star formation rates are also missing. 

Qualitatively, NIHAO galaxies reproduce the broad observed MaNGA spiral galaxy properties. 
However, some discrepancies remain.
In the next section, we quantify our model-observation comparisons by measuring ODR slopes and intrinsic scatters for the 12 scaling relations presented in \Fig{manga_nihao_compare}.

\subsection{Quantitative Comparisons} \label{sec:scatter}

\begin{figure*}
    \centering
    \includegraphics[width=\linewidth]{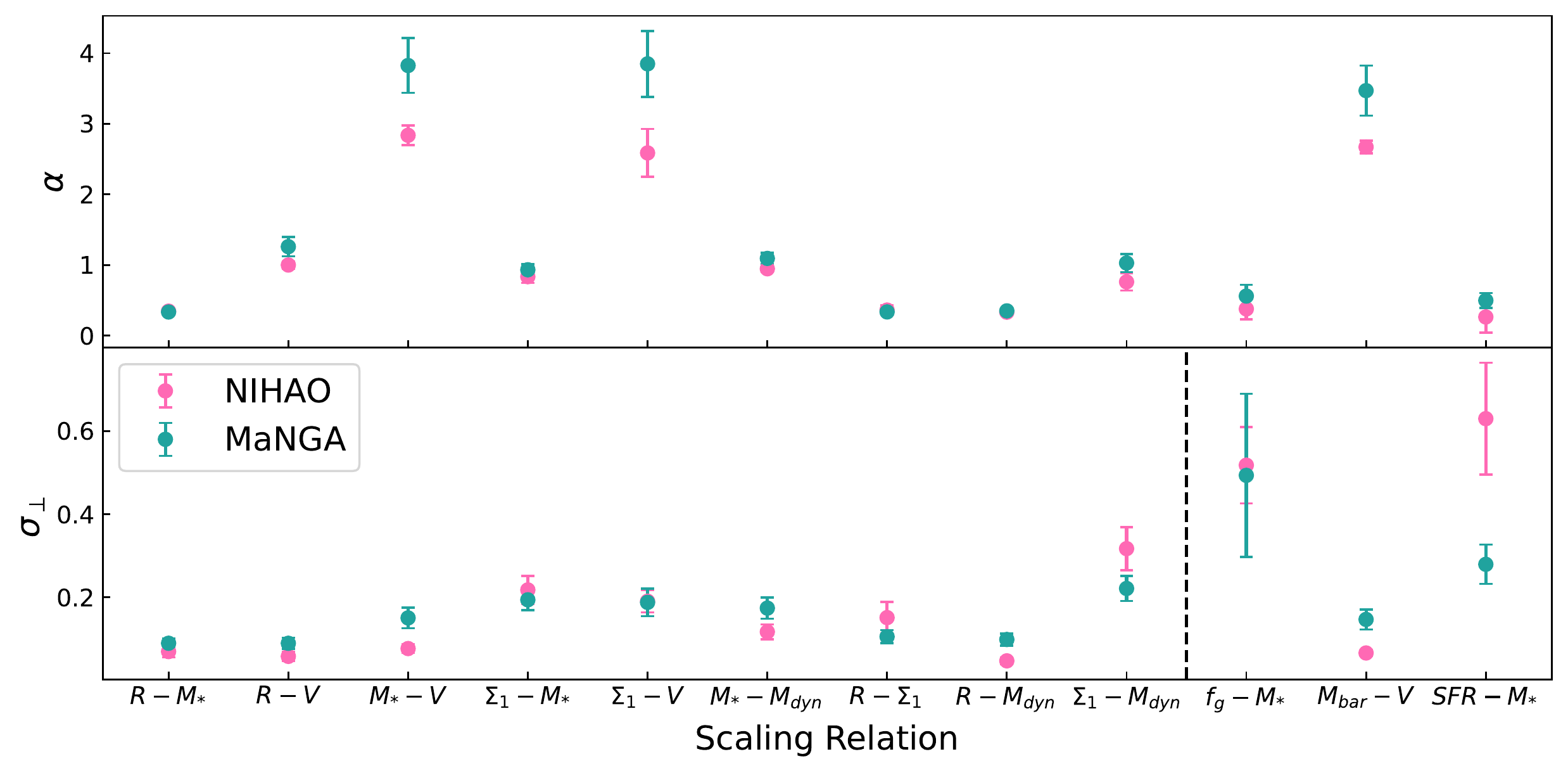}
    \caption{Comparison of the slopes and orthogonal scatters for the MaNGA and NIHAO scaling relations presented in \Fig{manga_nihao_compare}. 
    The top panel shows the slopes, $\alpha$, of the scaling relations, while the bottom panel gives the orthogonal scatters, $\sigma_{\perp}$.
    All properties for both observed and simulated galaxies are calculated at a radius which corresponds to a stellar surface density of 10\,M$_{\rm \odot}\,$pc$^{-2}$ (with the exception of the gas mass and SFR which are obtained using Pipe3D).
    The pink and teal dots represent simulated NIHAO and observed fits, respectively.
    The x-axis shows each scaling relation.
    The black dashed line in the bottom panel shows scaling relations based on galaxy properties from Pipe3D. 
    }
    \label{fig:slope_scatter_compare}
\end{figure*}

With the observed MaNGA sample being two orders of magnitude larger than the simulated NIHAO sample, a fair quantitative comparison of the respective scaling relations requires similar numbers of observed and simulated galaxies.
The selection of the MaNGA galaxies is performed randomly to match the size of the smaller NIHAO samples while keeping the NIHAO stellar mass distribution.  
Specific scaling relation fits are then carried out. 
The random selection of MaNGA sub-samples and measurements of the scaling relation fits were carried out 500 times; the slope and scatter and their corresponding errors were calculated as the median and standard deviations of the distribution.
It is noted that the fit parameters (slope, zero-point, and scatter) for the NIHAO-sized MaNGA sub-samples differ only marginally $(< 1\sigma)$ from the fit parameters based on the complete sample.
These results are presented in \Fig{slope_scatter_compare}.
However, the dearth of data points in the both the observed and simulated samples thwarts the assessment of robust Bayesian intrinsic scatters for our scaling relation comparisons. 
For observed galaxy scaling relations, the Bayesian intrinsic scatters were calculated and presented in \Tab{manga_nihao_sr}. 

The results presented in \Fig{slope_scatter_compare} show decent agreement for slopes and scatter of scaling relations, though differences exist.
Agreements are found for the derived slopes of various scaling relations, including the $R-M_{*}$, $\Sigma_1-M_{*}$, $M_{*}-M_{\rm dyn}$, and $f_{\rm g}-M_{*}$ relations.
Stellar mass is a common denominator in these scaling relations.
Conversely, disagreements between observed and simulated slopes exist for relations involving velocity estimates (specifically STFR, $\Sigma_1-V$, and BTFR).
Disagreements for scaling relations involving velocity metrics are also reflective of the simulations inabililty to reproduce the diversity of observed RCs and central stellar densities \citep{Oman2015, Frosst2022}.
Indeed, NIHAO galaxies show excess halo expansion due to over-efficient stellar feedback which results in lower central densities and different RC slopes than observations \citep{Dutton2017, Frosst2022}. 
For STFR and BTFR, NIHAO galaxies agree with theoretical scaling relation slope expectations \citep{Courteau2007} but a $1\sigma$ disagreement with observations exists.
NIHAO galaxies have been shown to qualitatively reproduce observed slopes for the BTFR \citep{Dutton2017, McQuinn2022}; albeit with fewer galaxies.

Along with the slope of scaling relations, the scatter informs us about the cosmic variations within galaxy formation and evolution processes.
The bottom panel of \Fig{slope_scatter_compare} shows the observational and simulated orthogonal scatters for the MaNGA sample. 
All galaxy structural scaling relations presented in \Fig{slope_scatter_compare} show reasonable agreement between simulations and observations.
Roughly, $1\sigma$ disagreements exist for STF, BTF, and SFMS relations.
A source of discrepancies between MaNGA and NIHAO, and specifically for the SFMS, owes to the differences in measured or inferred SFRs. 
While both simulations and observations infer SFRs over the last 100\,Myr, SFR values retrieved from Pipe3D \citep{Sanchez2018} are measured within 1.5$R_{\rm e}$ (the extent of the MaNGA IFU) while NIHAO properties are measured within a spatial location corresponding to 10\,M$_{\odot}$\,pc$^{-2}$. 
Measurement differences also exist for the BTFR which involves gas masses (retrieved from Pipe3D); however, as gas contributions to the mass budget for massive LTGs are relatively small ($f_{\rm g}\sim 10\%$), the slope and scatter disagreements for our data-model comparison remain small.

In conclusion, while NIHAO structural scaling relations agree reasonably well with MaNGA observations, some differences are observed. 
The latter can be attributed to the lack of RC diversity \citep{Santos2018, Frosst2022} and lower central stellar densities as a result of the strong feedback prescriptions in NIHAO.
However, some of the differences (specifically scaling relations involving gas masses and SFR) stem from different measurement techniques. 
This is evidenced by the large scatters of the $f_{\rm g}-M_{*}$ and SFMS relations.

\begin{table*}
\begin{center}
\begin{tabular}{@{}lcccccccc@{}}
\toprule
\multicolumn{3}{c}{Sample}                                                              & \multicolumn{3}{c}{MaNGA}                                                                        & \multicolumn{3}{c}{NIHAO}                           \\ \midrule
Scaling Relation             & $y$            & \multicolumn{1}{c|}{$x$}                & $\alpha$      & $\sigma_{\rm obs}$ & \multicolumn{1}{c|}{$\sigma_{i, \perp}$} & $\alpha$      & $\sigma_{i}$  & $\sigma_{i, \perp}$ \\ 
(1)                          & (2)            & \multicolumn{1}{c|}{(3)}                & (4)           & (5)                & \multicolumn{1}{c|}{(6)}                 & (7)           & (8)           & (9)                \\ \midrule
Project size-stellar mass    & $\log R$       & \multicolumn{1}{c|}{$\log M_*$}         & 0.37$\pm$0.01 & 0.09$\pm$0.01      & \multicolumn{1}{c|}{0.07$\pm$0.01}       & 0.35$\pm$0.01 & 0.08$\pm$0.01 & 0.08$\pm$0.01       \\
Project size-velocity        & $\log R$       & \multicolumn{1}{c|}{$\log V$}           & 1.33$\pm$0.02 & 0.14$\pm$0.01      & \multicolumn{1}{c|}{0.09$\pm$0.01}       & 0.99$\pm$0.05 & 0.08$\pm$0.01 & 0.06$\pm$0.01       \\
Stellar Tully-Fisher         & $\log M_*$       & \multicolumn{1}{c|}{$\log V$}         & 3.59$\pm$0.04 & 0.24$\pm$0.01      & \multicolumn{1}{c|}{0.08$\pm$0.01}       & 2.84$\pm$0.15 & 0.15$\pm$0.03 & 0.05$\pm$0.01       \\
$\Sigma_1$-stellar mass      & $\log\Sigma_1$ & \multicolumn{1}{c|}{$\log M_*$}         & 0.93$\pm$0.01 & 0.24$\pm$0.01      & \multicolumn{1}{c|}{0.17$\pm$0.01}       & 0.83$\pm$0.08 & 0.29$\pm$0.05 & 0.22$\pm$0.04       \\
$\Sigma_1$-velocity          & $\log\Sigma_1$       & \multicolumn{1}{c|}{$\log V$}     & 3.59$\pm$0.05 & 0.34$\pm$0.01      & \multicolumn{1}{c|}{0.11$\pm$0.01}       & 2.62$\pm$0.32 & 0.34$\pm$0.06 & 0.12$\pm$0.04       \\
Stellar mass-halo mass       & $\log M_*$     & \multicolumn{1}{c|}{$\log M_{\rm dyn}$} & 1.05$\pm$0.01 & 0.16$\pm$0.01      & \multicolumn{1}{c|}{0.13$\pm$0.01}       & 0.94$\pm$0.05 & 0.16$\pm$0.02 & 0.12$\pm$0.02       \\
$\Sigma_1$-size              & $\log\Sigma_1$ & \multicolumn{1}{c|}{$\log R$}           & 0.33$\pm$0.01 & 0.16$\pm$0.01      & \multicolumn{1}{c|}{0.15$\pm$0.01}       & 0.36$\pm$0.06 & 0.18$\pm$0.04 & 0.17$\pm$0.04       \\
Size-halo mass               & $\log R$       & \multicolumn{1}{c|}{$\log M_{\rm dyn}$} & 0.38$\pm$0.01 & 0.10$\pm$0.01      & \multicolumn{1}{c|}{0.09$\pm$0.01}       & 0.33$\pm$0.01 & 0.05$\pm$0.01 & 0.05$\pm$0.01       \\
$\Sigma_1$-halo mass         & $\log\Sigma_1$ & \multicolumn{1}{c|}{$\log M_{\rm dyn}$} & 0.97$\pm$0.02 & 0.32$\pm$0.01      & \multicolumn{1}{c|}{0.24$\pm$0.01}       & 0.75$\pm$0.12 & 0.42$\pm$0.06 & 0.34$\pm$0.05       \\
Gas mass-stellar mass$^*$        & $\log f_{g}$   & \multicolumn{1}{c|}{$\log M_*$}         & -0.65$\pm$0.03 & 0.39$\pm$0.01      & \multicolumn{1}{c|}{--}       & -0.37$\pm$0.14 & 0.42$\pm$0.11 & 0.39$\pm$0.13       \\
Baryonic Tully Fisher$^*$        & $\log M_{\rm bar}$       & \multicolumn{1}{c|}{$\log V$} & 3.32$\pm$0.04 & 0.24$\pm$0.01      & \multicolumn{1}{c|}{--}       & 2.65$\pm$0.08 & 0.13$\pm$0.02 & 0.05$\pm$0.01       \\
Star formation main sequence$^*$ & $\log SFR$     & $\log M_*$                              & 0.72$\pm$0.02 & 0.41$\pm$0.02      & --                           & 0.26$\pm$0.22 & 0.68$\pm$0.14 & 0.66$\pm$0.14       \\ \bottomrule
\end{tabular}
\end{center}
\caption{ODR linear fits for various scaling relations using MaNGA and NIHAO galaxies evaluated at the radius corresponding to a stellar surface density of 10\,M$_{\rm \odot}\,$pc$^{-2}$. 
Column (1) shows the scaling relation, and columns (2-3) give the variables for the corresponding relation. 
Columns (4-6) show the slope ($\alpha$), observed scatter ($\sigma_{\rm obs}$), and orthogonal Bayesian intrinsic scatter ($\sigma_{i, \perp} = \sigma_{i}/\sqrt{1+\alpha^2})$ of each scaling relation for the observed MaNGA sample. Likewise, columns (7-9) give the slope, scatter ($\sigma_i$) and orthogonal scatter ($\sigma_{i, \perp}$) for the NIHAO galaxies. 
The errors in each fit parameters are calculated using 1000 bootstrap runs.
The `` -- '' indicate the scaling relations for which Bayesian intrinsic scatter could not be calculated.}
\label{tab:manga_nihao_sr}
\end{table*}

\section{Spatially-Resolved Properties} \label{sec:char_radius}

We now take advantage of our spatially-resolved data to address the notion of 
spatially-resolved scaling relations in the fundamental VRM$_*$ parameter space.
Once again galaxy scaling relations, along with the observational errors, reflect fluctuations in the different processes that contribute to the distribution of galaxy properties in the Universe. 

To infer the VRM$_*$ properties, the rotation curves and stellar mass profiles were interpolated (or in some cases extrapolated) at radii based on stellar mass surface density.
If a certain stellar mass density does not exist, the object is discarded; i.e. no extrapolation is performed.
The process is applied consistently between MaNGA and NIHAO objects.
Ideally, the proper analysis of spatially resolved galaxy scaling relations should involve a Bayesian intrinsic fit parameter to analyse the amplitude of such a slope/scatter curve as a function of radius.
For the computation of Bayesian intrinsic fit parameters, the variation of each galaxy structural property and its associated error with galactocentric radius must be known \citep{Stone2019, Stone2020}.
Major sources of error on most structural parameters, and their corresponding scaling relations, include mass-to-light ratios, distance, and intrinsic disk thickness.  
Some of these vary with radius (e.g., M*/L, disk thickness), others not (distance).  
However, in all cases, we assume the error profile to be constant with galactocentric radius. 
As a result, the presentation of such spatially resolved parameter would only alter the amplitude of the function and not the shape (which is of interest here).

\subsection{Slopes}

\begin{figure}
    \centering
    \includegraphics[width=\columnwidth]{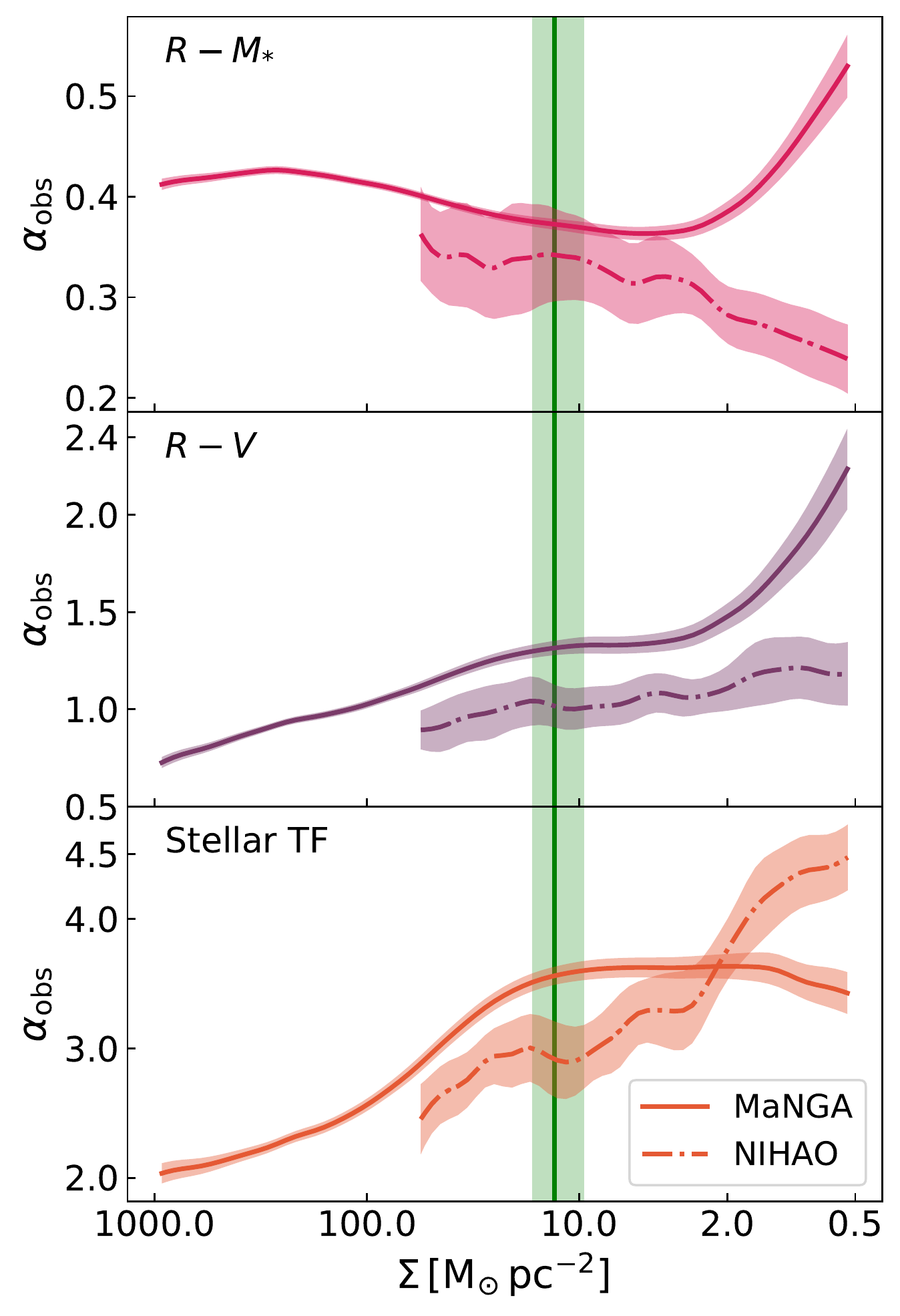}
    \caption{Variation of observed and simulated VRM$_*$ slopes as a function of radius scaled using stellar mass surface density. 
    The solid and dashed-dotted lines represent the MaNGA observations and NIHAO simulations respectively. 
    The green vertical line and shaded regions correspond to an isophotal level of 23.5\,mag\,arcsec$^{-2}$ in the \textit{z}-band for the MaNGA sample.}
    \label{fig:slope_variation}
\end{figure}

\Fig{slope_variation} shows the spatially-resolved slopes for the VRM$_*$ as a function of stellar mass surface density for the MaNGA and NIHAO galaxies. 
As expected, all three scaling relations for MaNGA and NIHAO galaxies show significant local (spatially-resolved) variations in galaxies.
The slope of the $R-M_*$ relation varies from $0.411\pm 0.003$ in the inner parts to the $0.544\pm 0.017$ in the outskirts. 
However, at $\Sigma_*\sim 6\,{\rm M_{\odot}\,pc^{-2}}$ the slope of $\sim$\,0.36 is closer to theoretical expectations, $R\propto M^{\sim 1/3}$ \citep{Courteau2007}.
In the outskirts of MaNGA galaxies ($\Sigma_*<2\,{\rm M_{\odot}\,pc^{-2}}$), the slope of the $R-M_{*}$ relation is less stable with fewer extended galaxies and larger size errors.
On the other hand, NIHAO simulations have the opposite trends for the spatially-resolved slopes for the $R-M_{*}$.
While agreement is found for $\Sigma_*\gtrsim2\,{\rm M_{\odot}\,pc^{-2}}$ for simulations and observations, differences are found at low stellar mass surface densities where scatter in our size metric is high.

The slope of the $R-V$ relation shows great variation, monotonically increasing from $\alpha_{\rm obs}\sim 0.7$ in the inner parts of galaxies to $\alpha_{\rm obs}\sim 2.4$ in the outskirts. 
The $R-V$ relation has a constant slope of $\sim\,$1.2 in regions corresponding to $\Sigma_*\sim 2-10\,{\rm M_{\odot}\,pc^{-2}}$ which represents the plateau in galaxy RCs. 
In very low stellar surface density regimes, the $R-V$ slope increases greatly and shows larger errors. 
This once again corresponds to the large diversity in low stellar mass surface densities in the outer regions of galaxies due to sporadic star formation.
The NIHAO simulated galaxies show a roughly constant slope for the $R-V$ relation as a function of stellar mass surface density.
While in tension with the observations, the constant slope for the NIHAO $R-V$ relation agrees with theoretical expectations, $R\propto V^{\sim 1.1}$ \citep{Courteau2007}.

Finally for the STFR, the observed slopes increase with radius to reach an agreement with the theoretical expectations of $\alpha \sim 3-4$ for values of $\Sigma_*\sim 10\,{\rm M_{\odot}\,pc^{-2}}$ or less. 
Beyond this characteristic stellar surface density, the observed STFR slope remains constant.
This is explained by a combination of the cumulative stellar mass profile and RC flattening leading to a constant contribution of baryons. 
NIHAO simulations show slightly increasing slopes for the STFR up to a threshold surface density of $\Sigma_*\sim 2\,{\rm M_{\odot}\,pc^{-2}}$, beyond which the slope sharply rises. 
Such behaviour is indicative of rising RC and/or stellar mass profiles 
(\Figs{nihao_sig_prof}{nihao_rc}) for NIHAO galaxies \citep[also][]{Frosst2022}.

In conclusion, similar to the spatially-resolved scatters, the range of VRM$_*$ slopes changes greatly with the adopted size metrics (calculated here using stellar mass surface density).
The simulated NIHAO galaxies result in smaller VRM$_*$ slopes than inferred observationally; 
the disagreement is at the $2-\sigma$ level and so represents only weak tension.
Similar to the scatter, the spatially-resolved slopes of scaling relations as a function radius can effectively constrain realistic numerical models of galaxies. 

\subsection{Scatters} 

The variation of scatter as a function of spatial position within galaxies traces the degree to which the universe creates a uniform (or variable) set of mass (baryon and dark matter) distributions.
The variation of scatter as a function of spatial positions within galaxies is a powerful tool to connect astrophysics at the different scales of galaxy formation. 
To understand the variation of the scatter for a scaling relation, we study the variation of the ``normalized scatter'' ($\sigma_{\perp}/\sigma_{\parallel}$) as a function of a physically-motivated size metric (measured here using projected stellar mass surface densities).
If a scaling relation is constructed using galaxy properties $X$ and $Y$ with a slope, $\alpha$, and observed scatter, $\sigma_{\rm obs}$, then the orthogonal scatter, $\sigma_{\perp} = \sigma_{\rm obs}/\sqrt{1+\alpha^2}$, informs us about the scatter normal to that scaling relation.
The choice of orthogonal scatter enables a uniform comparison of spatially-resolved scatters while taking into account the variations in the slope as well.
$\sigma_{\parallel}$ represents the scatter parallel to a scaling relation; 
in other words, it corresponds to the dynamical range of a scaling relation in a two-dimensional space.
Mathematically, the parallel scatter is the standard deviation calculated as: 
\begin{equation}
    \sigma_{\parallel} = std\Bigg[\frac{-X}{\sqrt{\alpha^2+1}}+\frac{-\alpha Y}{\sqrt{\alpha^2+1}}\Bigg].
    \label{eq:sig_p}
\end{equation}
The ratio, $\sigma_{\perp}/\sigma_{\parallel}$, is therefore, able to represent how informative a scaling relation is. 
Since for the same orthogonal scatter, if a relation was spread over a greater dynamic range ($\sigma_{\parallel}$) then there is more meaningful diversity in that population. 
This can be seen by considering the opposite limit of $\sigma_{\perp} = \sigma_{\parallel}$ in which the data is an uninformative circle. 

\begin{figure*}
    \centering
    \includegraphics[width=\columnwidth]{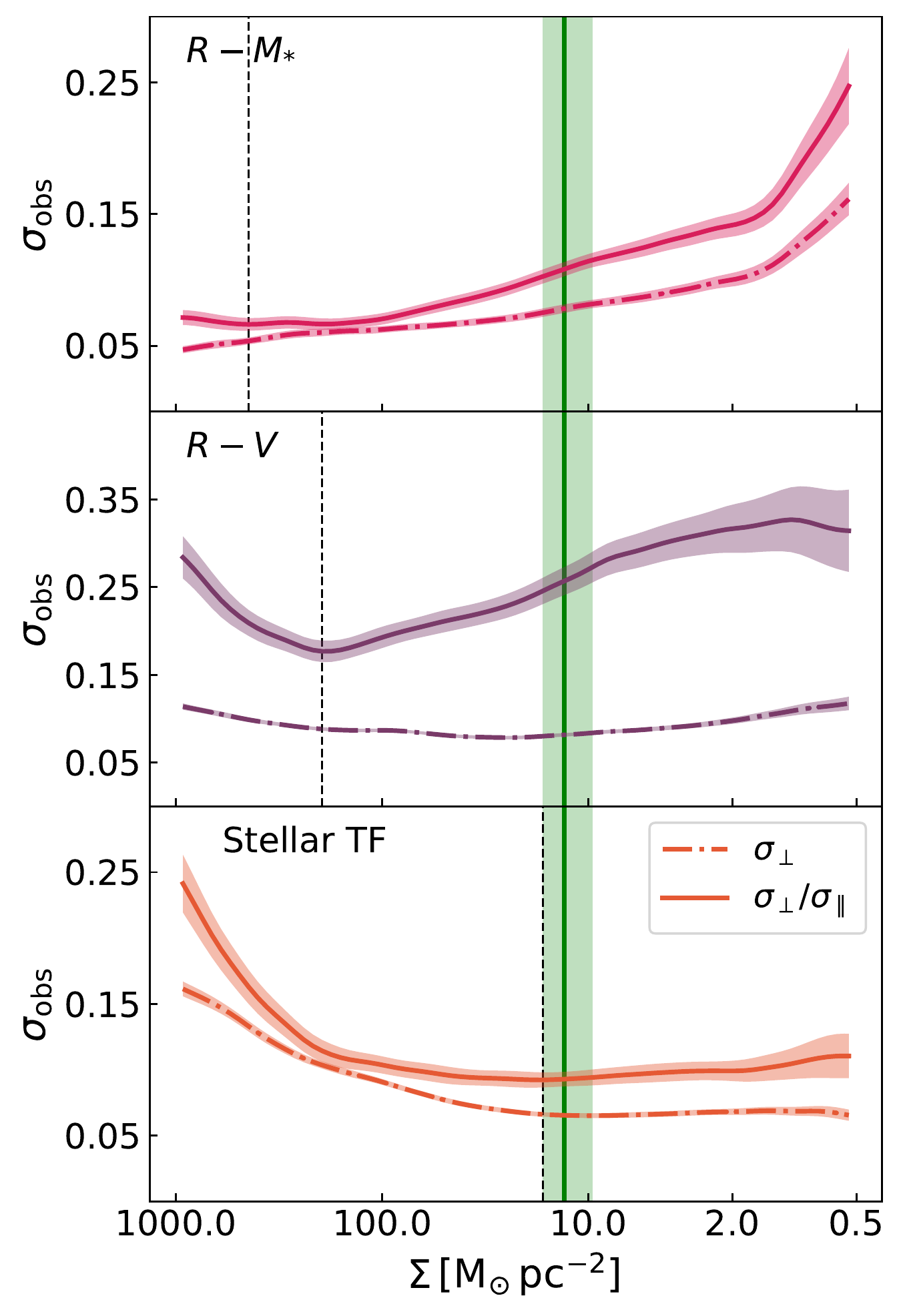}
    \includegraphics[width=\columnwidth]{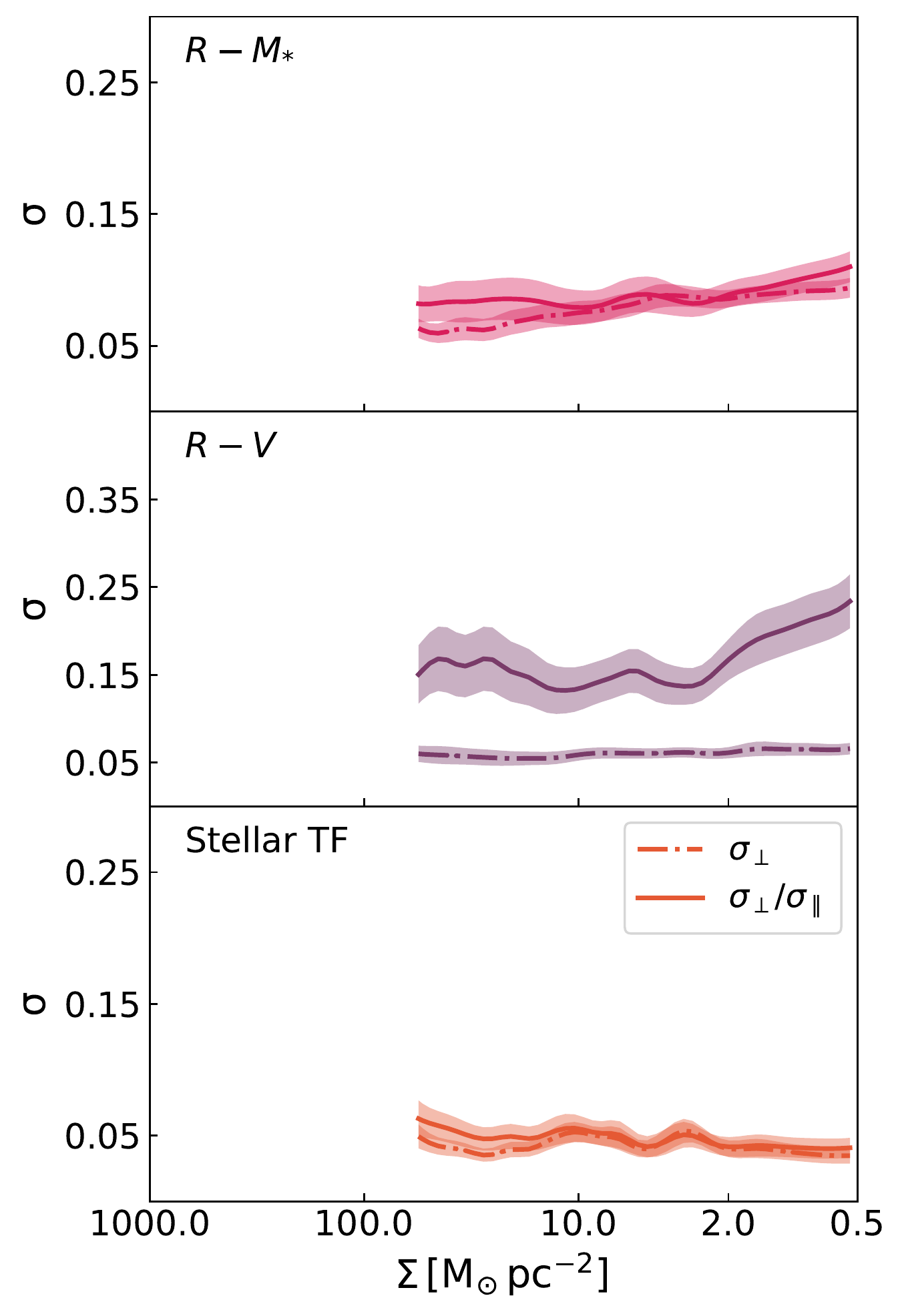}
    \caption{Variation of the scatter as a function of stellar surface density for the VRM$_*$ scaling relations.
    (Left) Showing observational MaNGA data. 
    (Right) Showing the simulated NIHAO data.
    The dashed-dotted and solid lines represent the orthogonal scatter ($\sigma_{\perp}$)and the normalized scatter ($\sigma_{\perp}/\sigma_{\parallel}$), respectively. 
    The shaded regions show the error calculated for the scatter using 1000 bootstrap runs. 
    For clarity, the shaded region represents twice the error. 
    For the left column (all three windows), the dashed black vertical line shows the isophotal radius where $\sigma_{\perp}/\sigma_{\parallel}$ is minimized. 
    The green vertical line and shaded regions correspond to an isophotal level of 23.5\,mag\,arcsec$^{-2}$ in the \textit{z}-band for the MaNGA sample.
    The x-axis baseline is the same in all windows to facilitate a direct comparison.}
    \label{fig:scatter_variation}
\end{figure*}

\Fig{scatter_variation} shows variations of the scatter metric as a function of stellar mass surface density for the VRM$_*$ scaling relations for MaNGA and NIHAO galaxies.
The green vertical line represents the normalized and orthogonal scatter calculated for a stellar surface density range of $9.5-16.7\,{\rm M_{\odot}\,pc^{-2}}$, which itself corresponds to an isophotal level of approximately 23.5\,mag\,arcsec$^{-2}$ in the \textit{z}-band for MaNGA LTGs.
Independent of the scatter metric, the VRM$_*$ scaling relations show a significant variation of $\sim$0.2\,dex for different sub-galactic regions within the MaNGA sample. 
Thus, regardless of the size metric, the scatter variations of VRM$_*$ scaling relations can inform us about the radial dependence of processes controlling the evolution of galaxy structures.
Analysing spatially-resolved scatter for scaling relations (e.g., for the radial acceleration relation \citep{Lelli2017} and the SFMS \citep{Wuyts2013, Cano2016, Delgado2016, Wang2017, Hall2018, Ellison2018}) provides a more complete understanding of the physics shaping galaxy properties in the Universe.
As with the spatially resolved SFMS, this spatially resolved scatter can allow us to connect the physics on the local scale to the global astrophysics.

The top panel in the left-hand column of \Fig{scatter_variation} shows variations of the scatter metrics for the observed MaNGA size-mass relation.
The dashed vertical black line represents the density where the lowest normalized scatter is recorded.
The minimized normalized scatter shows the region where a particular scaling relation provides the most information; this is a region where the dynamic range ($\sigma_{\parallel}$) is maximized and the orthogonal scatter ($\sigma_{\perp}$) is minimized.
For the $R-M_{*}$ relation, the minimum normalized scatter is found at $\Sigma_{*}\sim444\,{\rm M_{\odot}\,pc^{-2}}$ or $19.6\pm0.3\,{\rm mag\,arcsec^{-2}}$ in the \textit{z}-band.
The minimum density and SB level are found in the inner parts of the LTGs where the bulge component dominates the light and mass budget. 
From this analysis, it is evident that the stellar contents of galaxy bulges are quite uniform while disk components show greater diversity.
The orthogonal scatter (dashed-dotted line) for the $R-M_{*}$ relation decreases with increasing stellar surface densities.
The larger stellar surface densities probe inner/bulge dominated regions within LTGs; these regions behave similarly to the tighter (smaller scatter) size-mass projections for ETGs \citep{Lange2015, Trujillo2020, Arora2021}.
The smaller scatter is the result of bulge growth via repeated wet/dry mergers \citep{Shen2003, Company2013}.

The correlation between galaxy sizes and velocity informs us about the growth of angular momentum in galaxies, via the connection between luminous and dark content in galaxies.
The lowest normalized scatter in the $R-V$ relation is found at $\Sigma_{*}\sim 196\,{\rm M_{\odot}\,pc^{-2}}$.
This corresponds to a \textit{z}-band isophotal level of $20.4\pm0.04\,{\rm mag\,arcsec^{-2}}$, 
which is typically found in the inner parts of LTGs.
The larger normalized scatter for higher stellar surface densities results from the wide variations in the rising parts of LTG RCs~\citep{Oman2015, Oman2019, Frosst2022}.

We contrast the diversity of inner RCs with the larger inner scatter of the size-velocity and stellar TFR.
The larger diversity of the low stellar surface density ($\Sigma_*<196\,{\rm M_{\odot}\,pc^{-2}}$) is the result of a broader distribution of galaxy sizes.
The quartile range of isophotal radius increases as lower stellar surface density are examined. 
This behaviour is similar to the $R-M_*$ relation where larger normalized scatters are obtained in the disk-dominated regions of LTGs.
The larger scatters on both sides of the minimum density are indeed a result of the parameter diversities for LTGs.
In the inner parts of galaxies, velocity metrics show significant diversity~\citep{Oman2015, Oman2019, Frosst2022}, while physically-motivated galaxy sizes (corresponding to stellar mass surface densities) show larger diversities in the galaxies outskirts.
Such diversity in physically-motivated galaxy sizes are a result sporadic star formation in the outer regions of galaxies \citep{Baras2018}. 

The combination of $R-M_*$ and $R-V$ relations also yields the STFR which directly relates the luminous and dark matter content in galaxies. 
The minimum scatter of the STFR is found at $\Sigma_*\sim 17{\rm M_{\odot}\,pc^{-2}}$, which corresponds to an isophotal level of $23.2\pm 0.7\,{\rm mag\,arcsec^{-2}}$ (in the $z$-band).
This result agrees with other findings that the Tully-Fisher relation has its scatter minimized at an isophotal size of 23.5\,${\rm mag\,arcsec^{-2}}$ (in the $i$-band)~\citep{Giovanelli1994,Courteau1996,Hall2012}.
Slight differences between specific isophotal levels emerge from our measure of scatter based on normalized scatter.
The larger scatters for ($\Sigma_*>17\,{\rm M_{\odot}\,pc^{-2}}$) are due to the inner shapes of the stellar mass and velocity profiles.
The larger diversity in the shapes of the stellar mass and velocity profiles leads to a large inner normalized and orthogonal scatter.
The smaller, near constant, normalized and orthogonal scatter are a result of the flat nature of stellar mass profiles and rotation curves in the outer/disk dominated regions of the MaNGA galaxies.

The study of spatially-resolved scatters (normalized and orthogonal) has shown that the inner scatters of the combined VRM$_*$ scaling relations are dominated by the great diversity of galaxy RCs; only reflected in the scaling relation involving a velocity metric.
Conversely, scaling relations built on parameters measured in the galaxy's outskirts are controlled by the non-uniformity in stellar surface density, likely driven by stochastic star formation and feedback.
It is evident from the left-hand panel of \Fig{scatter_variation} that the scatter varies significantly as a function of radius.
Therefore, for a more complete understanding of the physics of galaxy scaling relations, we caution against the use of scatter based on a single radial metric.
Given the availability of IFU surveys, such as MaNGA\citep{Bundy2015, Wake2017}, CALIFA \citep{Walcher2014}, and SAMI \citep{sami}; and deep imaging such as provided by DESI \citep{Dey2019} and LSST \citep{Ivezic2019}; spatially-resolved scaling relations must be investigated in order to achieve a thorough understanding of structure evolution on local and global scales.

We note that the interpretation of the spatially-resolved scatter for VRM$_*$ scaling relations in NIHAO simulations depends greatly on sample definition and the size metric. 
For example, the trends in \Fig{scatter_variation} can differ for dwarf only samples and if size definitions are based on fractions of total light (see \photopaper for more details).
However, with MaNGA's target selection being restricted to $\log (M_{*} / {\rm M_{\odot}) > 8.5}$, our conclusions apply mostly to massive LTGs.

The right-hand column in \Fig{scatter_variation} shows spatially-resolved scatters for the VRM$_*$ scaling relations using the NIHAO simulations.
Given the (small) central densities observed in \Fig{nihao_sig_prof} for NIHAO galaxies, we cannot achieve the large stellar surface densities seen for MaNGA galaxies in \Fig{scatter_variation}.
We can only study resolved scatters for $\Sigma_{*}<50\,{\rm M_{\odot}\,pc^{-2}}$ using NIHAO galaxies.
The x-axis baseline for the left and right panels in \Fig{scatter_variation} were made equal to enable uniform model-observation comparisons.
The lower number of zoom-in simulations slightly increases the random errors within our simulated fits for the VRM$_*$ relations. 

Qualitatively, for all the VRM$_*$ relations, NIHAO and MaNGA galaxies have similar spatially resolved scatter trends. 
For example, the spatially resolved scatter (orthogonal and normalized) for NIHAO and observed galaxies increases with decreasing stellar surface density.
Similarly, the STFR normalized spatially resolved scatters for both MaNGA and NIHAO stay roughly constant for $\Sigma_{*}\leq 50\,{\rm M_{\odot}\,pc^{-2}}$.
The same is broadly true for the $R-V$ relations.

However, though these scatter trends are similar, their absolute values differ ($\sim\,$0.05 dex) due to the contribution of observed errors in MaNGA galaxies.
Furthermore, NIHAO simulations struggle to produce the high stellar surface densities found in the observed MaNGA galaxies. 
While we have demonstrated the usefulness of spatially-resolved scatters, a larger, morphologically-diverse observed sample would enable a less-biased study. 

\section{Conclusions} \label{sec:conclusions}

We have presented a dynamical catalogue of galaxy RCs for MaNGA LTGs, as a complement to the extensive photometric and environmental catalogue of \photopaper.
Based on these data and the NIHAO zoom-in simulations, an extensive comparison between galaxy observations and simulations of galaxies was established to highlight similarities and differences.
 
The MaNGA \ha velocity maps were fitted with an inclined rotating disk model assuming a hyperbolic tangent model for the circular velocities as a function galactocentric radius.
The fitting procedure allowed us to extract inclination-corrected RCs for MaNGA galaxies which provided us with accurate dynamical properties.
The tanh velocities fit the MaNGA \ha velocity cubes very well, with an average error of $\sim$7$\,{\rm km\, s^{-1}}$.
However, errors as high as $\sim$15-20$\,{\rm km\, s^{-1}}$ were recorded due to non-circular flows and PSF effects in the galaxy's central parts.
While non-circular motions affect our study only weakly, their investigations would greatly benefit the comprehensive and rigorous analysis of galaxy velocity fields. 

The combination of RCs with robust photometry for the MaNGA galaxy reveals the full extent of the MaNGA spectral data.
With robust photometry based on deep imaging ($\mu_z \sim$26\,${\rm mag\,arcsec^{-2}}$), we establish that the primary+ (secondary) MaNGA samples extend out to $1.4\pm0.4$ ($1.9\pm0.4$) R$_e$.
In general, our DESI photometry collects 0.3 mag more light than the NSA SDSS photometry which explains the lesser extent of the MaNGA velocity fields in units of effective radii.
On the other hand, the use of isophotal radius is independent of total light and provides a more accurate extent for the MaNGA IFU.
In units of R$_{23.5}$ ($z$-band), the primary+ (secondary) MaNGA spectroscopic data extend out to $0.8\pm0.2$ ($1.0\pm0.2$).

The ten RCs overlapping the MaNGA and PROBES surveys (the former using \ha velocity maps, the latter long-slit \ha RCs) show excellent agreement, with average differences no greater than $\sim $20\,km\,s$^{-1}$.
MaNGA velocity maps and extracted RCs are indeed well suited for analyzing the dynamics of a galaxy's interiors. 
The RCs for MaNGA allow for the construction of spatially-resolved structural scaling relations. ODR fit parameters for the VRM$_*$ scaling relations (measured at isophotal levels) are provided in \tab{manga_fits} and \Fig{manga_vrm}.
The VRM$_*$ scaling relations from MaNGA agree well with other similar relations presented in the literature \citep[see][for more comparisons]{Stone2020}, including studies using MaNGA galaxies.

For uniform comparisons between simulations and observations, the  metric at which structural parameters are measured is no longer isophotal but rather corresponding to a stellar mass surface density threshold. 
A stellar mass density of 10\,M$_{\odot}$\,pc$^{-2}$ was chosen for this comparison as it closely relates to the common isophotal R$_{23.5}$.
The versatility and richness of the MaNGA data enable comparisons of observations and simulations through multiple scaling relations. 
Using dynamical information, DESI photometry (\photopaper), gas and SF properties, we construct 12 galaxy scaling relations to compare observed MaNGA and simulated NIHAO galaxies. 
NIHAO simulations broadly reproduce the observed LTG populations ($
\log (M_{*} / {\rm M_{\odot}) > 8.5}$) seen in the MaNGA data. 
Qualitatively, most NIHAO galaxies lie within the observed 1$\sigma$ region (see \Fig{manga_nihao_compare}). 
Good matches are indeed found for the $R-M_{*}$, $\Sigma_1-M_{*}$, $M_{*}-M_{\rm dyn}$, and $f_{\rm g}-M_{*}$ relations. 
The simultaneous comparison using multiple scaling relations is beneficial in constraining various astrophysical phenomena (star formations, dynamics, stellar populations, etc.)

However, as a result of their complex selection function, NIHAO simulations also lack galaxies with high stellar mass, high circular velocity, and high star formation rates.
The largest discrepancies between NIHAO and MaNGA are found for scaling relations involving velocity metrics. 
Both of these are controlled by the feedback prescription and halo response to feedback \citep{McCarthy2012, Dutton2016}. 
For scaling relations involving gas masses, NIHAO galaxies produce a larger gas fraction than observed in low stellar mass galaxies ($\sim$0.3\,dex) due to excess cooling.
However, part of that discrepancy may be due heterogeneous measurements for gas masses.
Any excess gas does not result in high SFR for NIHAO galaxies; rather, the NIHAO SFMS show broad agreement with the observed one \citep[see also][]{Blank2021}.

The spatially-resolved scatter for VRM$_*$ scaling relations for MaNGA and NIHAO systems is also studied.
The scatter of scaling relations encodes astrophysical processes that cause cosmic variation in galaxy formation and evolution.
We study the variations of the normalized scatter, which provides information about the orthogonal dispersion as well as the dynamical range of a scaling relation, as a function of stellar mass surface densities. 
In the inner parts, the larger scatter of the $R-V$ and STFR are dominated by the diversity of galaxy RCs \citep{Frosst2022}.
Conversely, for the $R-M_*$ relation, the normalized scatter in the inner regions remains low likely due to uniform bulge populations and formation mechanisms \citep{Courteau1996b,Kormendy2004}.
The scatter for the VRM$_*$ scaling relations at large galactocentric radii is dominated by the diverse stellar mass surface densities. 
The effect of the diverse stellar mass surface density is most evident in the $R-M_*$ relation where the scatters, both orthogonal and normalized, continue to increase with decreasing stellar surface density.
The larger scatter in stellar mass surface density could result from sporadic star formation at larger galactic radii due to mergers and interactions.

We also compare the spatially-resolved scatters of observed MaNGA and simulated NIHAO galaxies. 
For the surface densities available in NIHAO, the simulated $R-M_*$ relation is broadly consistent with MaNGA data. 
Furthermore, both the $R-V$ and STFR relations are broadly consistent with a constant scatter as a function of stellar surface density, albeit for lower surface stellar density. 
The spatially-resolved scatters also show that NIHAO fails to produce the diverse range of observed RCs \citep{Frosst2022}. 
Most conspicuously, NIHAO simulations do not produce the high stellar masses and central densities that are seen in MaNGA data.
Other similar numerical simulations fail in similar fashion 
\citep{Oman2015, Oman2019, Roper2022}.

While NIHAO can broadly reproduce properties of the observed LTG galaxy population, some discrepancies remain such as those pertaining to central stellar densities and the diversity of RCs due to AGN feedback. 
With the spatially-resolved data presented here, we may now constrain simulations at all possible physically-motivated radii which represent different sub-galactic environments.
Not only should simulations be able to reproduce global galaxy scaling relations (size-mass, SFMS, TFR), they should also match the observed spatially-resolved properties and relations (such as the diversity of inner RC slopes, spatially-resolved SFMS, RAR, metallicity gradients, etc.)
These detailed comparisons bring together the complex interplay between local and global astrophysical processes. 

\section*{Acknowledgements}

NA, SC, and CS are grateful to the Natural Sciences and Engineering Research Council of Canada, the Ontario Government, and Queen's University for generous support through various scholarships and grants.
This material is based upon work supported by Tamkeen under the New York University Abu Dhabi Research Institute grant CAP$^3$.
NA also acknowledges support from the Arthur B. McDonald Canadian Astroparticle Physics Research Institute.
The research was performed using the {\scriptsize PYNBODY} package \citep{Pontzen2013}, S{\scriptsize CI}P{\scriptsize Y} \citep{scipy}, and N{\scriptsize UM}P{\scriptsize Y} \citep{numpy} and used {\scriptsize MATPLOTLIB} \citep{matplotlib} for all graphical representation.
The authors gratefully acknowledge the Gauss Centre for Supercomputing e.V. (www.gauss-centre.eu) for funding this project by providing computing time on the GCS Supercomputer SuperMUC at Leibniz Supercomputing Centre (www.lrz.de).
This research was carried out on the High Performance Computing resources at New York University Abu Dhabi. 
We greatly appreciate the contributions of all these computing allocations.

Funding for the Sloan Digital Sky Survey IV has been provided by the Alfred P. Sloan Foundation, the U.S. Department of Energy Office of Science, and the Participating Institutions.

SDSS-IV acknowledges support and resources from the Center for High Performance Computing at the University of Utah. The SDSS website is https://www.sdss.org.

SDSS-IV is managed by the Astrophysical Research Consortium for the Participating Institutions of the SDSS Collaboration including the Brazilian Participation Group, the Carnegie Institution for Science, Carnegie Mellon University, Center for Astrophysics | Harvard \& Smithsonian, the Chilean Participation Group, the French Participation Group, Instituto de Astrofísica de Canarias, The Johns Hopkins University, Kavli Institute for the Physics and Mathematics of the Universe (IPMU) / University of Tokyo, the Korean Participation Group, Lawrence Berkeley National Laboratory, Leibniz Institut für Astrophysik Potsdam (AIP), Max-Planck-Institut für Astronomie (MPIA Heidelberg), Max-Planck-Institut für Astrophysik (MPA Garching), Max-Planck-Institut für Extraterrestrische Physik (MPE), National Astronomical Observatories of China, New Mexico State University, New York University, University of Notre Dame, Observatário Nacional / MCTI, The Ohio State University, Pennsylvania State University, Shanghai Astronomical Observatory, United Kingdom Participation Group, Universidad Nacional Autónoma de México University of Arizona, University of Colorado Boulder, University of Oxford, University of Portsmouth, University of Utah, University of Virginia, University of Washington, University of Wisconsin, Vanderbilt University, and Yale University.

\section*{Data Availability}
All observed data for this paper have been incorporated into the online supplementary material.
The simulated data underlying this article will be shared upon reasonable request to the corresponding author.



\bibliographystyle{mnras}
\bibliography{reference} 



\appendix

\section{The Kinematic Catalogue}\label{sec:cat}

This section presents the headers for the kinematic catalogue based on the 2368 MaNGA LTGs used in this study (\Tab{manga_dyn_cat}).
The extracted fit parameters for our velocity model are provided in this supplementary material.
\begin{table*}
\begin{tabular}{@{}llcc@{}}
\toprule
Column Name & Description                                                    & Unit         & Data Type \\ 
(1)         & (2)                                                            & (3)          & (4)       \\ \midrule
MaNGA-ID    & MaNGA Identification                                           & ---          & string    \\
RA          & Right Ascension of the object (J2000)                          & $^{\circ}$   & float     \\
DEC         & Declination of the object (J2000)                              & $^{\circ}$   & float     \\
Z           & NSA or SDSS redshift                                           & ---          & float     \\
TType       & Morphological T-Type (from MDLM-VAC)                          & ---          & float     \\
XC          & X coordinate for the centre of the observed galaxy             & arcsec       & float     \\
dXC         & Error in XC                                                    & arcsec       & float     \\
YC          & Y coordinate for the centre of the observed galaxy             & arcsec       & float     \\
dYC         & Error in YC                                                    & arcsec       & float     \\
Vsys        & Systemic (heliocentric) velocity of the object                              & km\,s$^{-1}$ & float     \\
dVsys       & Error in V\_sys                                                & km\,s$^{-1}$ & float     \\
PA          & Fit position angle                                             & $^{\rm rad}$   & float     \\
dPA         & Error in PA                                                    & $^{\rm rad}$   & float     \\
I           & Fit inclination                                                & $^{\rm rad}$   & float     \\
dI          & Error in inclination                                           & $^{\rm rad}$   & float     \\
Vm          & Fit V$_{\rm max}$ for the tanh function (\Eq{V_tanh})          & km\,s$^{-1}$ & float     \\
dVm         & Error in Vm                                                    & km\,s$^{-1}$ & float     \\
Rt          & Fit R$_{\rm t}$ for the tanh function (\Eq{V_tanh})            & arcsec       & float     \\
dRt         & Error in rt                                                    & arcsec       & float     \\\bottomrule
\end{tabular}
\caption{Kinematic quantities for the MaNGA galaxies.
Columns (1), (2) and (3)  present the column name, description, units, and type for a given kinematic parameter, respectively. 
The table is found in the supplementary material in a comma separated values file format.}
\label{tab:manga_dyn_cat}
\end{table*}


\bsp	
\label{lastpage}
\end{document}